\documentclass[useAMS,usenatbib]{mn2e}
\usepackage{graphicx}
\usepackage{amssymb}
\usepackage{enumerate}

\newcommand{\apj}{ApJ}

\newcommand{\aap}{A\&A}

\newcommand{\mnras}{MNRAS}

\newcommand{\asprv}{ASPRv}
\newcommand{\pasj}{PASJ}
\newcommand{\ssr}{Space Sci. Rev.}
\newcommand{\msun}{M_\odot}

\begin{document}
\title[X-ray variability of SS\,433]
{X-ray variability of SS\,433: effects of the supercritical accretion disc}

\author[K. Atapin, S. Fabrika, A. Medvedev and A. Vinokurov]
{Kirill Atapin,$^1$\thanks{E-mail: atapin.kirill@gmail.com; fabrika@sao.ru}  
Sergei Fabrika,$^{2,3}$
Aleksei Medvedev$^{4,2}$ and Alexander Vinokurov$^2$ \\
$^1$  Sternberg Astronomical Institute, Moscow State University, Universitetsky pr., 13, Moscow 119991, Russia\\
$^2$ Special Astrophysical Observatory, Nizhnij Arkhyz 369167, Russia \\
$^3$ Kazan Federal University, Kremlevskaya 18, Kazan 420008, Russia \\
$^4$ Faculty of Mathematics and Mechanics, Saint-Petersburg State University, Universitetsky pr., 28, Saint-Petersburg 198504, Russia}
 
\pagerange{\pageref{firstpage}--\pageref{lastpage}}
\pubyear{2013}
\date{\today}

\label{firstpage} 
\maketitle


\begin{abstract}
We study a stochastic variability of SS\,433 in the $10^{-4} - 5\times 10^{-2}$\,Hz frequency range based on RXTE data, and on simultaneous observations with RXTE and optical telescopes. We find that the cross-correlation functions and power spectra depends drastically on the precession phase of the supercritical accretion disc. 
When the wind funnel of the disc is maximally open to the observer, a flat part 
emerges in the power spectrum; a break is observed at the frequency $1.7\times10^{-3}$~Hz, with a power-law index $\beta \approx 1.67$ at higher frequencies. The soft emission forming mostly in the jets, lags behind the hard and optical emission. When the observer does not see the funnel and jets (the `edge-on' disc), the power spectrum is described by a single power-law with $\beta \approx 1.34$ and no correlations between X-ray ranges are detected. We investigated two mechanisms to explain the observed variability at the open disc phase, 1) reflection of radiation at the funnel wall (X-rays and optical) and 2) the gas cooling in the jets (X-rays only). 
The X-ray variability is determined by the contribution of both mechanisms, however the contribution of the jets 
is much higher. We found that the funnel size is $(2-2.5)\times10^{12}$~cm, and the opening angle is $\vartheta_f\sim 50^\circ$. X-ray jets may consist of three fractions with different densities: $8\times10^{13}$, $3\times10^{13}$ and $5\times10^{11}$ cm$^{-3}$, with most of the jet's mass falling within the latter fraction. We suppose that revealed flat part in the power spectrum may be related to an abrupt change in the disc structure and viscous time-scale at the spherization radius, because the accretion disc becomes thick at this radius, $h/r \sim 1$. The extent of the flat spectrum depends on the variation of viscosity at the spherization radius. 
\end{abstract}
 

\section{Introduction}
SS\,433 is the only known superaccretor in the Galaxy (see \citep{Fabrika2004} for a review). The binary system consists of a compact relativistic component, most probably a black hole \citep{Kubota2010}, and a massive star filling its Roche lobe. The supercritical accretion with a mass accretion rate of $\sim 10^{-4} M_{\odot}$ yr$^{-1}$ $\sim 300\dot{M}_{Edd}$, where $\dot{M}_{Edd}$ is the Eddington (critical) accretion rate, is realized in the system. The observed relativistic jets ($v_j\approx0.26c$) precess with a period of $P_{pr}\approx162$ days; the supercritical accretion disc is the source of these jets. SS\,433 is an eclipsing binary with an orbital period of $P_{orb}\approx 13.08$ days. The occultation of the relativistic component by the donor star occurs at the orbital phase $\varphi=0$. 

The structure of supercritical accretion discs was first described by \cite{ShakSun1973}. 
According to their concepts, if the matter inflow rate at the outer boundary of the disc exceeds a critical value, the disc should have a typical size-scale $r_{sp}$, below which the supercritical properties of the disc begin to manifest themselves. This size is called the spherization radius and depends only on the accretion rate. For SS\,433, $r_{sp}\sim10^9$\,cm. Below the spherization radius, the energy release in the disc reaches values at which the radiation pressure exceeds the gravity. A powerful outflow of matter in the form of jets and optically thick wind takes place in the supercritical region of the disc. This pattern is well confirmed by radiation-hydrodynamic simulations \citep{Ohsuga2005,Ohsuga2011,Okuda2009}. 

Within the framework of the \cite{ShakSun1973} model, where the wind is formed below the spherization radius, we can assume that the wind in SS\,433 has the shape of a hollow cone and forms a wind funnel (hereafter we refer to it as funnel). The study of eclipse depths in different energy ranges \citep{Cherepashchuk2005} showed that the size of the outer boundary of the funnel (or thick disc) is approximately equal to the size of the donor-star and amounts to $\sim10^{12}$\,cm. Depending on the precession phase $\psi$ an observer can see different depths inside the funnel. At the phase $\psi=0$ the disc is most open to the observer, and the angle between the funnel axis (and jet) and the line of sight reaches the  maximum value $\theta\approx57\degr$ \citep{Fabrika2004}. At those moments the optical spectrum is described by a blackbody with a temperature of $\sim 50000-70000$\,K \citep{Dolan1997}. At the phases $\psi\approx0.34,0.66$, the disc is oriented 'edge-on', and the observer should not be able to 
see the funnel at all.  \\

The X-ray luminosity of SS\,433 is $L_X\sim10^{36}$\,erg/s \citep{MedvFabr2010}, which is several orders of magnitude lower than its bolometric (mostly UV) luminosity $L_{bol}\sim10^{40}$\,erg/s \citep{Cherepashchuk2002}. The brightest object in the system is the supercritical accretion disc. Obviously, all the energy is initially released in the X-ray range. However, it thermalizes in the powerful wind of the supercritical disc and emerges as UV radiation. It was assumed for a long time that all the observed X-ray radiation originates in the jets. In the papers by \cite{Brinkmann1988} (EXOSAT), \cite{Kotani1996} (ASCA), \cite*{Marshall2002} (Chandra), and \cite{Filippova2006} (RXTE), the authors developed a `standard' cooling jet model and found the jet parameters. In particular, \cite{Marshall2002} obtained the following parameters: jet base temperature $T\sim 1.1\times10^8$\,K and opening angle $\vartheta_j\approx1.2\degr$.

The opening of the X-ray jets is equal to that of the optical jets~\citep{BorisFabr1987}. The typical radiative time-scale of X-ray jets is about one hundred seconds, whereas the radiative time for optical jets is 1--3 days \citep{Fabrika2004}. The opening angle of the X-ray jets is determined by the speed of sound in the place where they emerge from the funnel and begin to cool, $\sin{\theta_j}\simeq c_s/v_j$ \citep{Marshall2002}. The jet temperature subsequently drops, and the jet opening 'freezes' and changes no longer. 

An analysis of the new data from the XMM-Newton observatory showed that the standard jet model cannot explain the observed X-ray spectrum, and that an additional hard component in required \citep*{Brinkmann2005}. \cite{MedvFabr2010} analysed the XMM-Newton spectra (in the 0.2--12\,keV range) in more detail and isolated three components.  
(i)~Jet emission. The jet continuum emission dominates in the 1.5--5\,keV range; there is also a significant contribution in the iron and nickel lines in the vicinity of 7--8\,keV. 
(ii)~Reflected emission. It is assumed that the funnel walls can 'see' directly the bottom of the funnel and the hard radiation of the central engine. They reflect this radiation outwards. The reflected radiation dominates in the range of $>7$\,keV.
(iii)~The soft ($<1.5$ keV) thermal component~--- presumably the inherent radiation of the funnel walls. 

\cite{Revn2004} proposed an idea that if different emission components of SS\,433 originate in different spatially separated regions, the cross-correlation functions (CCFs) should exhibit a shift caused by the time lag between these components. \cite{Revn2004} and \cite{Burenin2011} investigated the correlations between X-rays (3--20\,keV) and the optical range. They suggested that the X-ray emission is formed in the jets, whereas optical emission is formed at the outer parts of the funnel's wall due to the thermal reprocessing of hard radiation, which is formed in the inner parts of the supercritical accretion disc and ends up on the funnel walls. If this is the case, one can expect a shift to exist between the X-ray emission of the jet (2--5\,keV) and the funnel ($>7$\,keV). The correlation between different X-ray ranges has not been studied so far, although, presently, a lot of new data are emerging, allowing such a study to be carried out.  

Another approach to the study of funnel parameters is investigating the aperiodic variability of SS\,433. \cite{Revn2006} constructed broad-band power density spectra (PDS) in the $10^{-8}$--$10^{-2}$\,Hz frequency range based on the data of optical, X-ray (EXOSAT/ME and RXTE/ASM) and radio observations. These authors suggested the idea that at the frequencies higher than $\sim 10^{-2}$\,Hz, PDS should become steeper due to the smearing out of variability in the funnel, and the position of the break in the power spectrum can be used to estimate the funnel size. PDS of SS\,433 in the visible were studied in the frequency range of $10^{-4}$~Hz and higher \citep{Burenin2011}. 

Thus, the nature of rapid variability, as well as the form of the observed CCFs and PDS should depend on the properties of the funnel and relativistic jets. At present, there are enough optical and X-ray (from the RXTE observatory) data accumulated to perform a more detailed and complete analysis of these unique structures in the accretion disc of SS\,433. The available data cover a full range of precession phases and have high signal-to-noise ratios, which allows to study the dependence of the form of the CCFs and PDS on disc orientation, model the observed PDS in detail, and estimate the funnel parameters. 

In section 2 we describe the selection criteria for observations and the reduction algorithm. In section 3 we present the power spectra for different precession phases. In section 4 we discuss the analysis of the correlation of two X-ray bands (in which the jet and the funnel presumably emit) with each other and with the optical.

\section{Observations}

\begin{table*}
\begin{minipage}{138mm}
  \caption{RXTE/PCA observations arranged by group and exposure. The first column gives the number of the group of observations; next, the observation ID, the numbers of the detectors used to analyse a given row, the date, $\psi$~--- precession phase, $\varphi$~--- orbital phase, $\theta$~--- angle between the line of sight and the funnel axis (jet) in degrees, $T_{obs}$~--- duration of the observation in ks, $T_{exp}$~--- useful exposure time after GTI filtering in ks, $R_{net}$ and $R_{bkg}$~--- the net and the background count rate per PCU in the 2--20\,keV range.}
  \label{tab:data_xray}
  \begin{tabular}{clcccccccc} \hline\hline
 Group & Obs. ID & PCU &Date & $\psi$ & $\varphi$ & $\theta$ & ${T_{obs}}$ &${T_{exp}}$ &$R_{net}(R_{bkg})$  \\\hline
  & 90401-01-01-00$^\dagger$ & 2,3 & 2004-03-13 & 0.98 & 0.486 & 60 & 25.9 & 15.4 & 44(9) \\ 
  & 90401-01-01-03$^\dagger$ & 2,3 & 2004-03-12 & 0.98 & 0.416 & 58 & 8.9 & 6.5 & 47(9) \\ 
  & 90401-01-01-02 & 1,2,3 & 2004-03-14 & 0.99 & 0.562 & 60 & 3.2 & 3.2 & 43(9) \\ 
  & 90401-01-03-02 & 2,3 & 2004-03-27 & 0.07 & 0.560 & 61 & 3.2 & 3.2 & 39(9) \\ 
I & 90401-01-04-01 & 2   & 2004-08-22 & 0.98 & 0.848 & 56 & 2.9 & 2.9 & 37(9) \\ 
  & 91103-01-01-00 & 2,4 & 2005-07-28 & 0.07 & 0.849 & 59 & 2.7 & 2.7 & 38(10) \\ 
  & 91103-01-06-01 & 2 & 2005-08-02 & 0.11 & 0.234 & 61 & 2.5 & 2.5 & 40(10) \\
  & 90401-01-01-01 & 2,3 & 2004-03-12 & 0.98 & 0.411 & 58 & 2.4 & 2.4 & 46(9) \\ 
  & 90401-01-03-01 & 2   & 2004-03-27 & 0.06 & 0.500 & 61 & 2.4 & 2.4 & 41(10) \\ 
  & 90401-01-03-00$^{*}$ & 2,3 & 2004-03-28 & 0.07 & 0.581 & 60 & 2.3 & 2.3 & 39(10) \\ \hline

   & 91092-01-02-00$^\dagger$ & 2 & 2005-08-06 & 0.13 & 0.515 & 66 & 25.9 & 15.1 & 37(10) \\
   & 91092-02-08-00 & 2 & 2005-08-16 & 0.19 & 0.282 & 71 & 19.9 & 8.8 & 32(9) \\ 
   & 10127-01-01-00$^\dagger$ & 0,1,2 & 1996-04-18 & 0.19 & 0.834 & 72 & 12.7 & 7.8 & 35(11) \\ 
   & 91092-02-06-00$^\dagger$ & 2 & 2005-08-15 & 0.18 & 0.186 & 69 & 8.6 & 6.1 & 33(9) \\ 
   & 91092-02-07-00$^\dagger$ & 2 & 2005-08-15 & 0.18 & 0.206 & 69 & 8.3 & 5.4 & 32(10) \\ 
II & 60058-01-15-00 & 2,4 & 2001-11-23 & 0.79 & 0.156 & 74 & 3.3 & 3.3 & 25(11) \\ 
   & 60058-01-17-00 & 1,2 & 2001-11-25 & 0.81 & 0.313 & 71 & 3.3 & 3.3 & 31(10) \\ 
   & 60058-01-10-00 & 2,3,4 & 2001-11-19 & 0.77 & 0.322 & 76 & 3.1 & 3.1 & 26(10)\\ 
   & 60058-01-16-00 & 2 & 2001-11-24 & 0.80 & 0.222 & 73 & 2.6 & 2.6 & 28(10) \\ \hline
 
    & 60058-01-02-00 & 2,3,4 & 2001-11-10 & 0.72 & 0.165 & 82 & 3.3 & 3.3 & 18(9) \\ 
    & 60058-01-05-00 & 2,3,4 & 2001-11-13 & 0.73 & 0.165 & 81 & 3.3 & 3.3 & 23(9) \\ 
    & 60058-01-06-00 & 2,3,4 & 2001-11-14 & 0.74 & 0.476 & 80 & 3.3 & 3.3 & 23(10) \\ 
    & 60058-01-07-00 & 2,3,4 & 2001-11-15 & 0.75 & 0.549 & 79 & 3.3 & 3.3 & 23(11) \\ 
III & 60058-01-04-00 & 2,3,4 & 2001-11-12 & 0.73 & 0.322 & 81 & 3.2 & 3.2 & 21(10) \\ 
    & 60058-01-08-00 & 2,3 & 2001-11-16 & 0.75 & 0.630 & 79 & 3.2 & 3.2 & 24(10) \\ 
    & 60058-01-03-00 & 2,3,4 & 2001-11-11 & 0.72 & 0.246 & 82 & 3.1 & 3.1 & 19(10) \\ 
    & 91103-01-10-00$^{*}$ & 2,3,4 & 2005-08-28 & 0.27 & 0.229 & 81 & 2.5 & 2.5 & 26(9) \\ \hline

   & 20102-02-01-06$^\dagger$ & 0,1,2,3,4 & 1998-03-06 & 0.43 & 0.337 & 83 & 28.6 & 16.2 & 22(11) \\ 
   & 30273-01-04-00$^\dagger$ & 0,1,2,3,4 & 1998-04-02 & 0.60 & 0.451 & 85 & 25.3 & 14.0 & 21(11)\\ 
   & 20102-02-01-00$^\dagger$ & 0,1,2,3,4 & 1998-03-06 & 0.43 & 0.386 & 83 & 26.4 & 13.7 & 21(12)\\ 
IV & 30273-01-02-01$^\dagger$ & 0,1,2,3,4 & 1998-03-30 & 0.58 & 0.229 & 84 & 23.9 & 11.1 & 21(11)\\ 
   & 30273-01-03-01$^\dagger$ & 0,1,2 & 1998-04-01 & 0.59 & 0.380 & 84 & 20.2 & 11.0 & 22(11)\\ 
   & 30273-01-03-00$^\dagger$ & 0,1,2,3 & 1998-03-31 & 0.58 & 0.303 & 84 & 19.7 & 10.9 & 22(10)\\ 
   & 30273-01-05-00$^\dagger$ & 0,1,2 & 1998-04-03 & 0.60 & 0.538 & 85 & 19.5 & 10.0 & 22(11)\\ \hline
  \end{tabular}
$^\dagger$The data sets were used to compute the power spectra\\
$^{*}$Joint X-ray and optical observations 
  \medskip 
  \end{minipage}
\end{table*}

In this study we analyse the data from the RXTE (Rossi X-ray Timing Explorer) archive, and also joint simultaneous RXTE and optical observations. We used only the data from the PCA (Proportional Counter Array) detector, because out of the three instruments onboard RXTE it has the maximum sensitivity. The archive contains more than a hundred observations in total. Most of them were made in 1998, 2004 and 2005.

Our aim was to study the nature of the variability of SS\,433 accretion disc structures as a function of precession phase. We used the most accurate of the currently available ephemerides to compute the precessional and orbital phases \citep{Goranskij2011}: the time of the maximum opening of the disc towards the observer ($\psi=0$) JD $2449998.0 + 162\fd278 \cdot E$; the time of eclipse of the relativistic star by the donor-star JD $2450023.746 + 13\fd08223\cdot E$. Since the eclipses by the donor-star are rather deep (although the amplitude of an eclipse depends strongly on wavelength, \citep{Cherepashchuk2005}), we selected non-eclipse data sets with the orbital phase $0.15\leq \varphi \leq 0.85$ for our analysis. To have a high signal-to-noise ratio, we decided to use data sets with exposure time of no less than 2\,ks. Ultimately, we selected 34 observations. 

\begin{table*}
\begin{minipage}{125mm}
  \caption{Joint observations with RXTE and optical telescopes, sorted by exposure time. The columns are denoted as in Table~\ref{tab:data_xray}, with the exception of $T_{ovl}$~--- the length of the overlapping parts of X-ray and optical light curves in ks; next, the optical telescope, band, and figure number.
} 
  \label{tab:data_xray+optics}
  \begin{tabular}{ccccccccc} \hline\hline
  Obs. ID              & Date       &$\psi$ & $\varphi$ & $\theta$ & ${T_{ovl}}$& Optical telescope & Band & Fig.\\ \hline
  91103-01-10-00 & 2005-08-28 & 0.27 & 0.229 & 81 & 2.5 & BTA & V & \ref{fig:xray+optics}a\\
  90401-01-03-00 & 2004-03-28 & 0.07 & 0.581 & 60 & 2.3 & RTT-150$^1$ & R & \ref{fig:xray+optics}b\\
  90401-01-02-01 & 2004-03-25 & 0.05 & 0.350 & 58 & 1.5 & RTT-150$^1$ & R & \ref{fig:xray+optics}c\\
  91103-01-05-01 & 2005-07-31 & 0.09 & 0.101 & 61 & 0.8 & RTT-150$^2$ & R & \ref{fig:xray+optics}d\\ \hline
  \end{tabular}
$^1$\cite{Revn2004}\\
$^2$\cite{Burenin2011} 
\end{minipage}
\end{table*}  

We divided all the observations into 4 groups depending on the precession phase (Table~\ref{tab:data_xray}). The depth to which an observer can see the inner wall of the funnel changes with the precession phase. The parameters of the kinematic model of SS\,433 are known with a high accuracy \citep{Eikenberry2001}: precession angle $\theta_{pr}=20\fdg92 \pm 0\fdg08$, orbit inclination $i=78\fdg05 \pm 0\fdg05$. This allows us to compute the angle $\theta$ between the line of sight and the funnel axis for each observation. The angle reaches its minimum value at precession phase $\psi=0$. In this phase, the observer can see deepest into the funnel. The nutational ''nodding'' of the jets and funnel with an amplitude of $\pm 2.8\degr$ also changes the value of the angle \citep{Fabrika2004}. To take into account the nutation, we used the $2450000.94 + 6\fd2877\cdot E$ \citep*{Goranskij1998} ephemerides, where the zero nutation phase corresponds to the maximum inclination of the funnel towards the observer. The Table~\ref{tab:data_xray} lists the value of angle $\theta$ corrected for nutation.  

We placed the observations in which the funnel is most open to the observer ($\theta<62\degr$) into the first group. Intermediate disc orientations are placed in the second and third groups. The fourth group contains the observations with the edge-on orientation of the disc ($83\degr \leq \theta \leq 90\degr$). In addition, the longest simultaneous X-ray and optical observations falls into the third group (Obs.\,Id 91103-01-10-00, $\theta\approx81\degr$, see below).

The observations were reduced using the HEASoft\,6.11 package. We used the data of `Good Xenon' mode. We tried to obtain the longest and most homogeneous light curves for each data set. In earlier observations, all 5 PCA detectors would usually operate. During the later observations, PCU1, PCU3 and PCU4 were periodically turned on and off due to a high voltage breakdowns. That is why we selected the Good-Time-Intervals (GTI) so as to use only the detectors that operated for a maximum length of time during the observations (Table~\ref{tab:data_xray}, column 3). In addition, the PCU0 detector that had lost the propane veto layer was excluded from consideration when reducing the observations made after the year 2000. Otherwise, standard parameters were used for creating the GTI. To calibrate the background, we used the L7/240 faint source model. The mean net count rate varies from about 20 to 45 counts/s/PCU depending on precession phase. The background count rate is about 10 for all observations (Table~\ref{tab:data_xray}).

We obtained the longest set of optical V-band observations (Table~\ref{tab:data_xray+optics}) with the 6-m BTA telescope of the Special Astrophysical Observatory (Russia) on August 28, 2005. A $2048\times 2048$ pixel EEV~CCD42-40 array was used as the detector. The magnitude of the target is $V\approx14^m$, and the flux measurement accuracy is 0.3\% at individual exposure time of 3~s. The full cycle of acquisition and readout (temporal resolution) was 10 seconds and varied insignificantly. During reduction, the light curve was interpolated to a uniform grid.

We added three other optical sets, simultaneous with X-ray observations, (Table~\ref{tab:data_xray+optics}) from \citet{Revn2004} and \citet{Burenin2011}. These R-band observations were performed with the 1.5-m Russian-Turkish Telescope (RTT150), T\"UBITAK National Observatory (TUG), Bakyrly mountain, Turkey. The $R$ magnitude is $\approx12^m$, the photometric accuracy is 2\% at individual exposure time of 1~s.

In addition to the reduction procedure described above, we performed time correction for four X-ray data sets in Table~\ref{tab:data_xray+optics}. The RXTE observatory clock uses the TT system (Terrestrial Time), whereas the optical observations are tied to the UTC system. At the time of observations, the difference between the two systems was equal to 64.184 seconds. This difference has been taken into account
(see ``Time Tutorial''\footnote{http://heasarc.nasa.gov/docs/xte/abc/time\_tutorial.html}).

\section{Power spectra}
\label{sec:pds}
The study of the power spectra of SS\,433 shows that its aperiodic variability is similar to the variability of the majority of Galactic X-ray sources and has a red noise nature \citep{Revn2004,Revn2006,Burenin2011}, i.e., its amplitude increases with increasing characteristic time-scales. The PDS averaged over the precession phases are well fitted by a power law $P \propto f^{-\alpha}$ with the exponent $1.5$ \citep{Revn2006}. However, a more detailed investigation of the power spectra, and in particular, their variation with the precession phase, allows one to find out more about the structure of the supercritical accretion disc of SS\,433. From the characteristic time-scales and the PDS slopes, we can estimate the size and opening angle of the wind funnel of the accretion disc.

\begin{figure*}
  \includegraphics[width=0.7\textwidth]{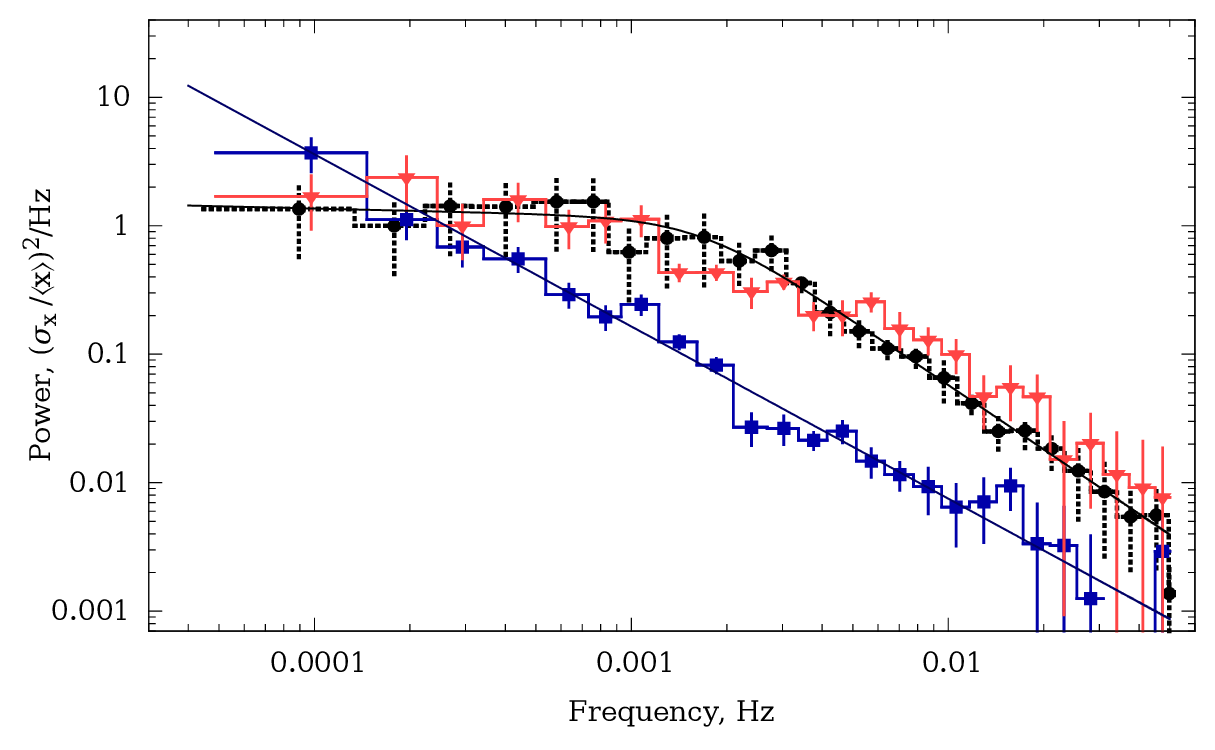}
\caption{Power spectra in the 2--20\,keV range for different precession phases (Table~\ref{tab:data_xray}). The circles and dotted line~--- the maximum opening of the disc towards the observer (group~I), triangles and grey (red) line~--- intermediate orientations (group~II),  squares and dark grey (blue) line~--- edge-on disc (group~IV). The solid lines show the model fits.  (A colour version of this figure is available in the online journal.)}
 
\label{fig:pds_data}
\end{figure*}

To compute the PDS we extracted light curves with a 10-second temporal resolution. Because the number of used detectors is varied, the light curves were normalized to this number. The light curves were then divided into intervals. Each interval was used to compute separate PDS, which were then averaged.

RXTE operates in low orbit with a period of about 1.5 hours, and all observations have significant gaps caused by the occultation of the object by the Earth, and also by the passage of RXTE through the South Atlantic Anomaly. Unfortunately, even in the best-case scenario, the total fraction of the gaps amounts to no less than 40\%. It was important to us to keep the low frequencies in the PDS, and therefore, we could not limit the length of the interval only to the unocculted ($\sim3000$\,s) parts of the light curves. We used then the 1024 bin ($\sim10$\,ks) intervals and omitted those where the fraction of gaps exceeds 50\%. To compute the PDS, we used the data sets that have at least one such interval left. They are marked in Table~\ref{tab:data_xray} by the ``$\dagger$'' sign. 

In Fig.~\ref{fig:pds_data} we show the 2--20\,keV X-ray PDS for the first, second, and fourth precession phase groups. The third group does not have observations long enough to compute the PDS. The PDS corresponding to the maximum opening of the funnel (group I) has an obvious break in the vicinity of $10^{-3}$\,Hz. As the funnel visibility conditions deteriorate for the observer (group II), the break becomes less conspicuous~--- the PDS gradually changes its slope in the vicinity of the break. However, at lower frequencies the PDS has the same slope as in the case of the maximally open disc. The power spectrum corresponding to the `edge-on' orientation of the disc (group IV) has no break at all. 

Fitting the fourth-group PDS with the power law $P \propto f^{-\alpha}$ gives $\alpha=1.34\pm0.19$. We fitted the first-group PDS with a broken power law
\begin{equation}
P \propto \frac{1}{f^{\beta_1} \sqrt{1+(f/f_{br})^{2\beta_2}}}
\label{eq:pds_knee}
\end{equation}
and obtained: $\beta_1=0.06\pm 0.09$, $\beta_2=1.61 \pm 0.14$, $f_{br}=(1.7\pm0.3)\times10^{-3}$. The spectral index asymptotically tends to $\beta_1$ at $f\ll f_{br}$, and to $\beta_1+\beta_2=1.67$ at $f\gg f_{br}$. At frequencies less than the break frequency $f_{br}$, the PDS is almost flat. This result does not agree with the index 1.5 found earlier by \cite{Revn2006}.

\begin{figure}\center
  \includegraphics[width=0.4\textwidth]{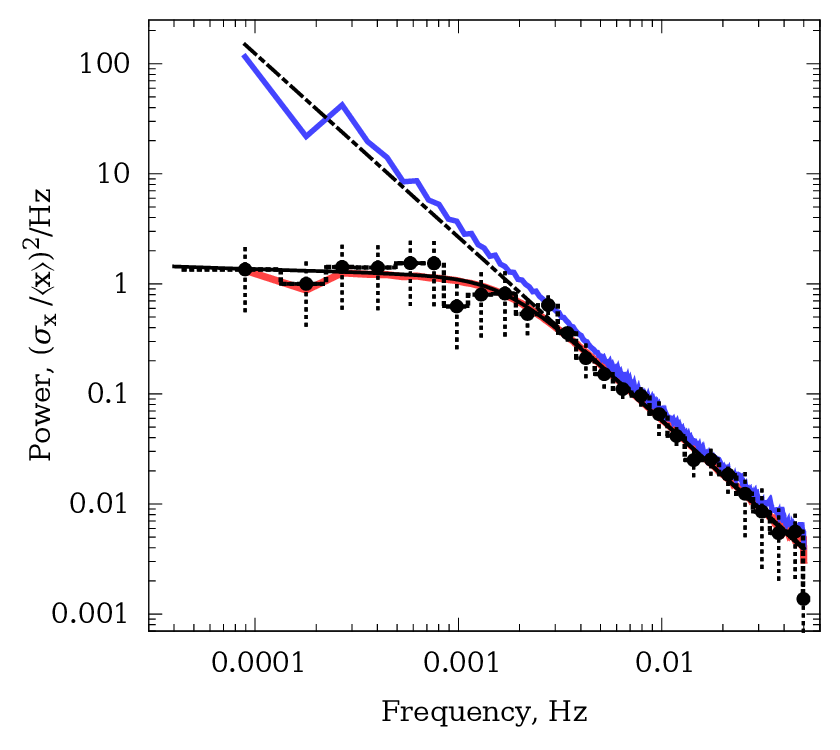}
\caption{Reality of the flat region in the power spectrum of SS\,433 in the phase of maximum funnel visibility. The circles are observed power spectrum (group I), the black solid and dashed-and-dotted lines~--- two original reference models. The same model spectra obtained with the allowance for gaps in observations are shown by the grey (red and blue) lines. The model with a flat region completely reproduces the observed PDS.  (A colour version of this figure is available in the online journal.)}
\label{fig:pds_flatnessproff}
\end{figure} 

Theoretically, the slope of the PDS may be distorted by the so-called power leakage \citep[Sect 6.1.3 and 7.5]{Priestley}, which may arise due to occasional gaps in the observations.
To check whether the flat region of the PDS is due to the influence of the gaps, we used the following Monte Carlo method. \cite{Timmer1995} describe an algorithm of generating synthetic light curves with a given power spectrum using the inverse Fourier transform. The measured power values $I(f_j)$ are scattered around the real $P(f_j)$, and obey the $\chi^2$ distribution with two degrees of freedom. Therefore, by assigning to each harmonic a random phase, we can generate synthetic light curves, i.e., realizations of a random process with the required statistical characteristics and PDS.

\begin{figure*}
  \includegraphics[width=0.4\textwidth]{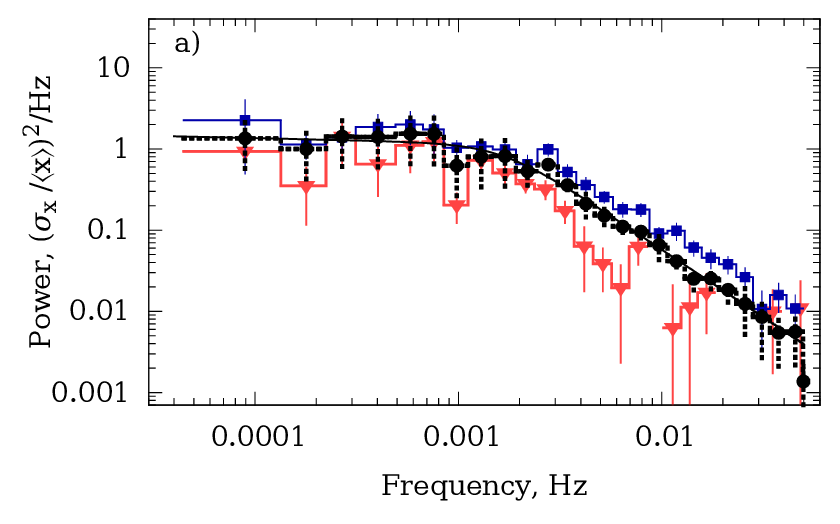}
  \includegraphics[width=0.4\textwidth]{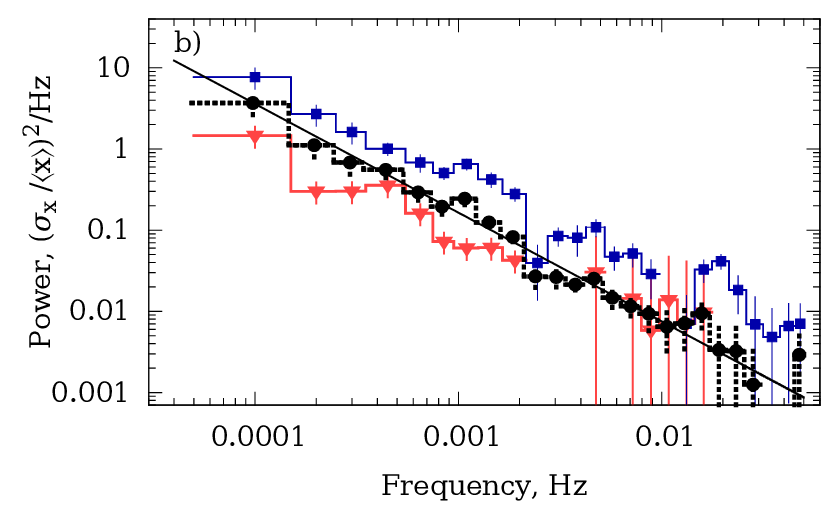}
\caption{Power spectra in different energy ranges:  8--20\,keV~--- squares and dark grey (blue) line, 2--5\,keV~--- triangles and grey (red) line and 2-20\,keV~--- circles and dotted line for maximally open funnel (a) and `edge-on' (b) orientations. (A colour version of this figure is available in the online journal.)}
\label{fig:pds_data_g1g4}
\end{figure*}

Using this algorithm, we specified a reference PDS, generated a large number ($\sim1000$) of synthetic light curves, superimposed a gap pattern present in the real observations, applied identical breaking into intervals and computed the synthetic light curves. By comparing the discrepancies between the reference PDS and the synthetic one computed using the Monte Carlo method, we can estimate the degree of distortion due to the gaps. As reference power spectra we used the one described by formula (\ref{eq:pds_knee}) and a single power-law PDS with the index 1.67, which well approximates the $f>f_{br}$ frequency range (Fig. \ref{fig:pds_flatnessproff}).

As is evident from Fig.~\ref{fig:pds_flatnessproff}, the presence of gaps does not lead to the appearance of a flat region in the single power-law PDS. In the broken power-law model a systematic underestimation of power is observed only at the frequency $\approx2\times10^{-4}$\,Hz. Otherwise there are no differences between the reference and synthetic PDS. Thus, the flat region at low frequencies in the PDS is not an artifact, and corresponds to the real processes in the SS\,433 accretion disc.
 
For the maximum opening of the funnel and edge-on phases, we separately show in Fig.~\ref{fig:pds_data_g1g4} PDS in the 2--5\,keV (the maximum jet emission) and 8--20\,keV (the maximum funnel emission) ranges. As is evident from the figure, there are no essential differences between the PDS in these two ranges. They both have the same shape typical of their groups. However, the PDS in the soft range is systematically lower. This indicates that the variability of emission in the soft and hard ranges has the same nature, but the amplitude of variability (in percent) is smaller in the soft range. 

\subsection{Funnel model}
\label{sec:funnel_model}
We found that the power spectrum corresponding to the open disc is practically flat at the low frequencies, and has a break at the frequency $f_{br}=1.7\times10^{-3}$\,Hz, above which the slope is substantially steep (1.67). The presence of a break in the PDS can be explained by the smearing out of variability in the funnel \citep{Revn2006}. It is assumed that the variable hard X-ray emission is generated in the innermost parts of the accretion disc, inaccessible for direct observations. An observer can see only the reflected emission. The signals reflected from the outer and inner parts of the funnel walls should be delayed relative to each other by the typical time $\tau \sim l_f/c$, where $l_f$ is the size of the funnel and $c$ is the speed of light. A superposition of these signals should result in the effective suppression of variability at times shorter than $\tau$ and the appearance of a break in the power spectrum. The particular shape of the observed PDS~--- the position of the break and the slope 
at high frequencies~--- should depend on the geometry of the funnel.  

We modeled the PDS shape depending on the funnel's geometry. In our model, the funnel of the supercritical accretion disc is represented by an opaque cone with an opening angle $\vartheta_f$ (the angle between the wall and the axis) and a funnel wall length $l_f$. The source of emission is point-like and located at the apex of the cone. We defined a coordinate system $(u,v)$ at the surface of the cone. We assumed that the illumination in each point of the cone is inversely proportional to the squared distance from the point source, and that the albedo is the same in all points. Then the radiation flux reflected in the direction of the observer by the surface element $dudv$ with radius vector $\vec{r}(u,v)$ is:
\begin{equation}\label{eq:funnelwall_flux}
F(u,v)\propto \frac{A\cos\gamma}{r^2},
\end{equation}
where A is the surface albedo and $\gamma(u,v)$ is the angle between the normal to the surface and the direction towards the observer. 

The smearing out of variability is described by the response function 
$R_{\vartheta_f,l_f,\theta}(t)$, which shows how the funnel with the given parameters reacts to an instant flare that occurred in the center of the accretion disc at time $t=0$:
\begin{equation}
R_{\vartheta_f,l_f,\theta}(t)=\frac{1}{\tilde{R}}\int\limits_{\Sigma(\vartheta_f,l_f,\theta)}F(u,v)\delta(t-\tau(u,v))dudv,
\end{equation}
where $\delta$ is the Dirac delta function. The function $\tau(u,v)\geq0$ describes the lag of the beam reflected by the element with the coordinates $(u,v)$. Integration is done over the inner cone surface visible to the observer. The delay function $\tau(u,v)$ and the visible surface area $\Sigma$ depend not only on the parameters of the cone, but also on its orientation with respect to the observer (on the angle $\theta$ between the line of sight and the funnel axis). In all our computations we adopted $\theta=60\degr$, which corresponds to the most long observation, which contributed the most to the power spectrum of group~I (Table~\ref{tab:data_xray}). The normalization $\tilde{R}$ was chosen in such a way so that the area underneath the response function would be equal to one. 

\begin{figure}\center
  \includegraphics[width=0.4\textwidth]{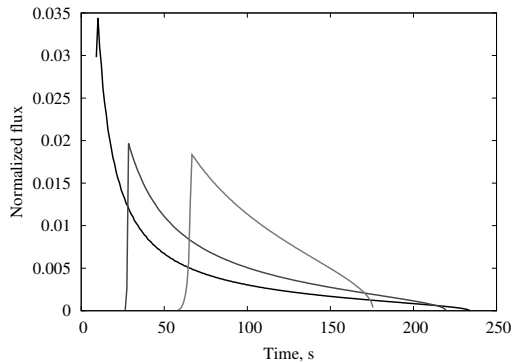}
\caption{Response functions of the funnel as a function of the opening angle. From left to right $\vartheta_f=55\degr$, 50\degr and 35\degr. The funnel length for all three models is equal to $l_f=5.1\times10^{12}$\,cm.} 
\label{fig:funnel_response}
\end{figure}

We computed the response functions for different values of the parameters $l_f$ and $\vartheta_f$ and found that the length of the funnel affects mainly the duration of the response, whereas the opening angle influences both the duration and steepness of the decline. In Fig.~\ref{fig:funnel_response} we show the response functions for different opening angles. All the functions have a characteristic shape with a sharp rise in the beginning and a long decline in the end. This is due to the fact that light reaches the deeper and, therefore, brighter regions of the funnel with the least delay. These deeper regions contribute the most to the response functions and correspond to the fast rise in the beginning. The peripheral areas of the funnel are reached by light after a longer delay. These areas are responsible for the formation of the long 'tail' in the response functions. 

\begin{figure}
  \includegraphics[width=0.48\textwidth]{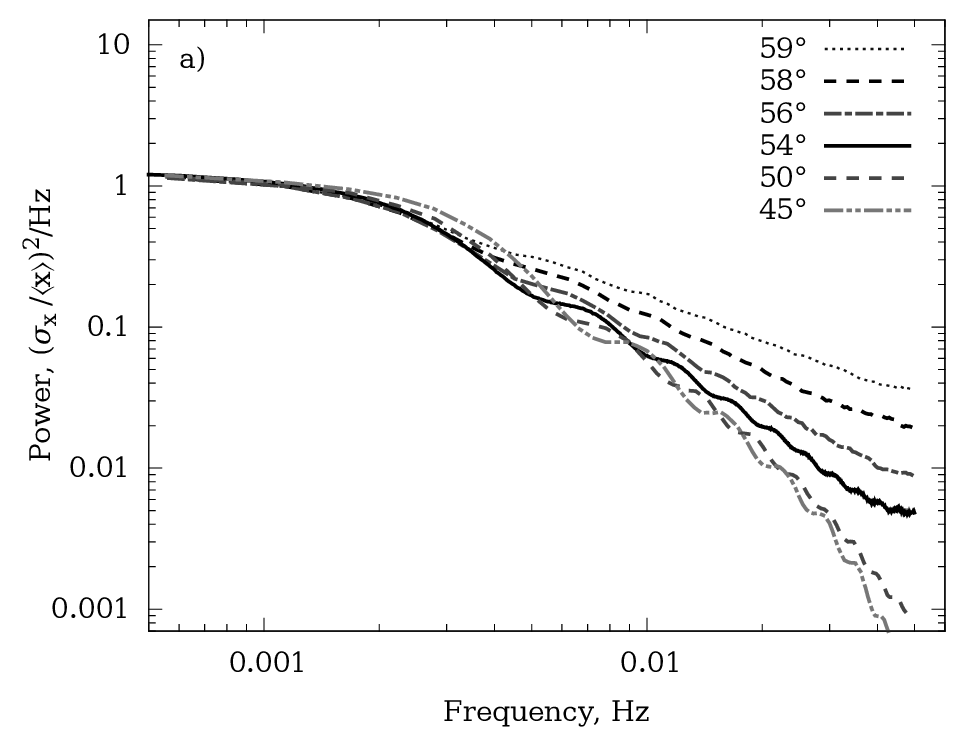}
  \includegraphics[width=0.48\textwidth]{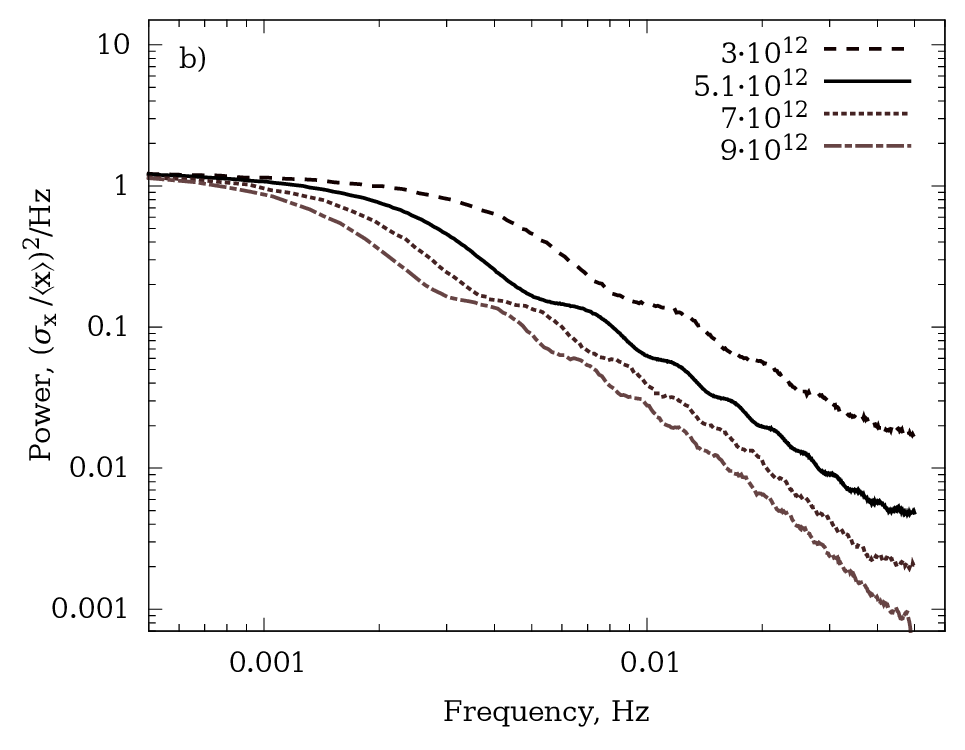}
   \includegraphics[width=0.48\textwidth]{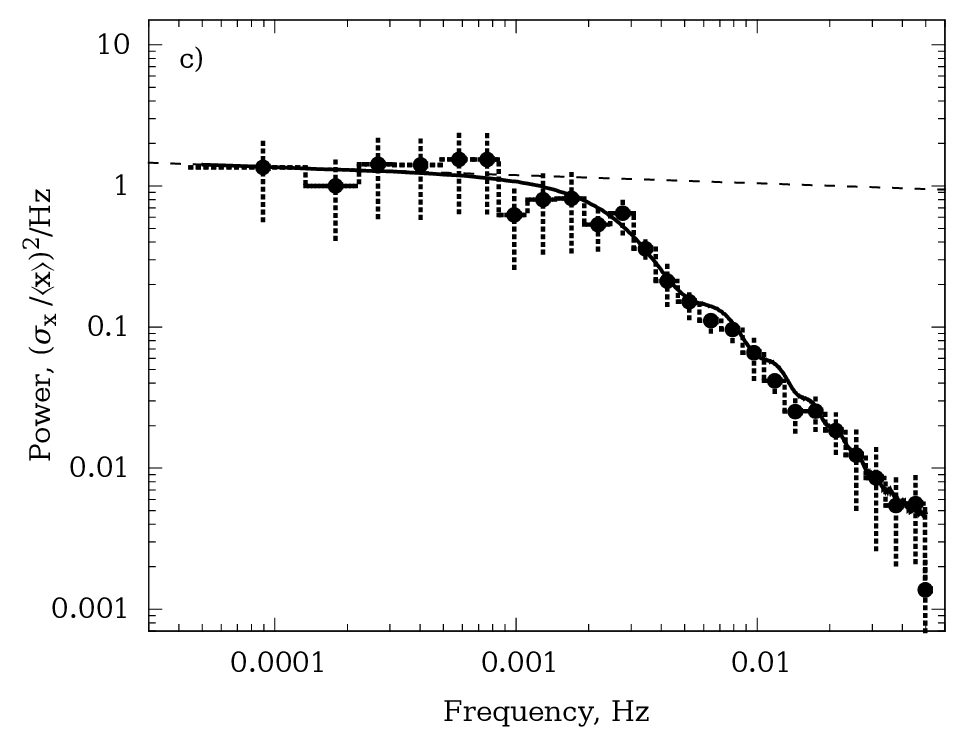}
\caption{Model power spectra for different values of the opening angle $\vartheta_f$ (a) ($l_f$ is adopted $5.1\times 10^{12}$~cm) and length $l_f$ (b) ($\vartheta_f=54\degr$) of the cone funnel. The best fit model is shown by the solid lines. (c)~--- the observed power spectrum (group I, maximally open disc) with the best fit model. The initial power spectrum of a point source with the power law $P\propto f^{-0.06}$ is shown by the dashed line.} 
\label{fig:pds_results}
\end{figure} 

As the opening angle increases, the more and more bright regions of the funnel become visible. That is why the maximum of the response functions is higher for large angles, and the decline is steeper. In the opening angle limit $\vartheta_f \rightarrow \theta$, degeneration occurs and the response function turns into a $\delta$-function, because the central source, which is much brighter than the funnel walls, becomes visible to the observer. 

Using the response functions we modeled the PDS. Since the effect of smearing out of variability in the funnel manifests itself at frequencies higher than the break frequency, we adopted that in the $f<f_{br}$ frequency range the slope of the PDS is $\beta\approx 0.06$ as it was found from observations. We suppose that the flat PDS is an intrinsic property of the innermost parts of SS\,433 accretion disc (see Sec.~\ref{sec:vis_time} below). Using the algorithm of \cite{Timmer1995}, we generated synthetic light curves corresponding to such flat power spectrum, convolved them with the response function of the funnel and constructed the model PDS. 
 
In Fig.~\ref{fig:pds_results}a,b we show the model power spectra for different opening angles and funnel lengths. As is evident from the figures, increasing the funnel length proportionally decreases the break frequency. The opening angle mainly determines the slope of the power spectrum at high frequencies. The slight influence of the angle on the position of the break is due to the change in the path difference between the beams reflected by the central and peripheral regions of the funnel.

As the opening angle increases, the power spectrum becomes flatter, and the sensitivity to the value of the angle increases. In the $\vartheta_f \rightarrow \theta$ limit, the slope of the power spectrum at high and low frequencies evens out, because the central source becomes visible. In this case, the model power spectrum corresponds to the flat power spectrum of the central source at all frequencies.  

The model power spectrum that best describes the observational data is shown in Fig.~\ref{fig:pds_results}c. The corresponding parameters are: opening of the funnel (the angle between the wall and the axis) $\vartheta_f\approx54\degr$, length of the funnel wall $l_f\approx5.1\times 10^{12}$ cm. The wave-like oscillations in the model PDS are due to the presence of sharp peaks in the response functions. Nonetheless, our simple model with only two parameters reproduces the shape of the observed PDS well. As is evident from Fig.~\ref{fig:pds_results}a, the model PDS are very sensitive to the values of the opening angle, and therefore the formal accuracy of angle determination is equal to a few degrees. The uncertainty in the funnel length is $\sim 20\%$.

In the modelling we assumed that the source of emission is in the center of the accretion disc and it is point-like. If the source has a size of the spherization radius $r_{sp} \sim 10^9$ cm, it is also much less than $l_f$. In this case formula (\ref{eq:funnelwall_flux}) must be multiplied by the cosine of the angle $\gamma_0$ between the beam incident on the wall and the normal to the surface. This angle is very close to $90\degr$ and $\cos{\gamma_0}\propto r_{sp}/r$. That is why the reflected emission flux becomes inversely proportional to the cube of the distance to the source $F(u,v)\propto r^{-3}$. 

Substituting the second power by the third influences the form of the response functions. Modelling of the power spectra with the new response functions yields the following model parameters: opening $\vartheta_f\approx45\degr$ and funnel wall length $l_f\approx7.3\times 10^{12}$ cm. The funnel length increases, whereas the opening decreases. Taking into account the other effects, e.g., the scattering of the radiation in the semi-transparent gas filling the funnel, should yield the same result (the cubic dependence). 

In the case of multiple scattering inside the funnel, the time of the propagation of variability is determined mostly by the last scattering in the outer parts of the funnel. This will not strongly change the result of our estimation of the opening and size of the funnel.

The wall length that we found apparently corresponds to the size at which the wall becomes too transparent and does not reflect quanta effectively. The radius of the wind photosphere $r_{ph}$ in SS\,433 may serve as an independent estimate of the funnel size. The wind that forms the funnel may probably emerges from the inner parts of the accretion disc \citep*{Vinokurov2013} in a certain range of angles $[\vartheta_f; \vartheta_f+\beta_w]$. The radius of the photosphere depends on the matter outflow rate $\dot{M}_0$, angle $\beta_w$ and wind velocity $v_w$:
\begin{equation}
r_{ph}=\frac{\dot{M}\sigma_T}{v_w\Omega \mu m_p },
\end{equation} 
where $\Omega=4\pi(\cos{\vartheta_f}-\cos{(\vartheta_f+\beta_w)})$ is the solid angle of wind propagation, $m_p$ is the mass of a proton, and $\sigma_T$ is the Thomson cross section. Using the values $\dot{M}_0\sim10^{-4}
\msun$~yr$^{-1}$ and $v_w\sim1000$ km/s 
\citep{Fabrika1997}, we obtain $r_{ph}\approx6.4\times 10^{12}$ cm for the angle $\beta_w=20\degr$. This value approximately corresponds to the funnel size found by us. From our modelling we estimate the funnel size along jet direction ($l_f\cos\vartheta_f$) as $\sim (2.5-3)\times 10^{12}$~cm what is about 1.5 times bigger than previous estimates \citep{Fabrika2004,Cherepashchuk2005}. 

Based on our modelling we can estimate the visible jet base radius $r_{j0}$. The orientation of SS\,433 is such that the inner parts of the funnel and the jet formation region are obscured by the wind. That is why the jet becomes visible to the observer only starting from a certain minimum distance from the black hole $r_{j0}$. This distance is included as a parameter in the standard cooling jet model and is determined from X-ray spectroscopic observations. \cite{MedvFabr2010} found $r_{j0}\approx2.8\times 10^{11}$\,cm from XMM-Newton spectra. Our estimate of this value, obtained based on the opening and the funnel length, is $r_{j0}=l_f\sin{(\theta-\vartheta_f)}/\sin{\theta}\approx 6.2\times 10^{11}$\,cm. Thus, the $r_{j0}$ obtained by us from the power spectra is approximately 2 times higher than the value obtained by \cite{MedvFabr2010}.

\section{Correlation functions}
The maximum of the SS\,433 jet emission falls within the 2--5\,keV range \citep{Kotani1996,Brinkmann2005,
MedvFabr2010}. Good RXTE sensitivity in the 2--20\,keV range allows studying both the variability of the jets and of the harder emission. It is quite probable that the harder emission ($>7$\,keV) is reflected \citep{MedvFabr2010} from the wind funnel walls. To study these two components of the X-ray spectrum~--- the jets and the reflected emission, we perform a cross-correlation analysis of the X-ray light curves in the ranges of 2--5\,keV and 8--20\,keV (hereafter we refer to them as ``soft'' and ``hard''). We have also studied the correlation between other sub-ranges, but have discovered that the effect is most conspicuous in the CCFs in the case of the energy ranges mentioned above. 

\begin{figure*}
  \includegraphics[width=0.81\textwidth]{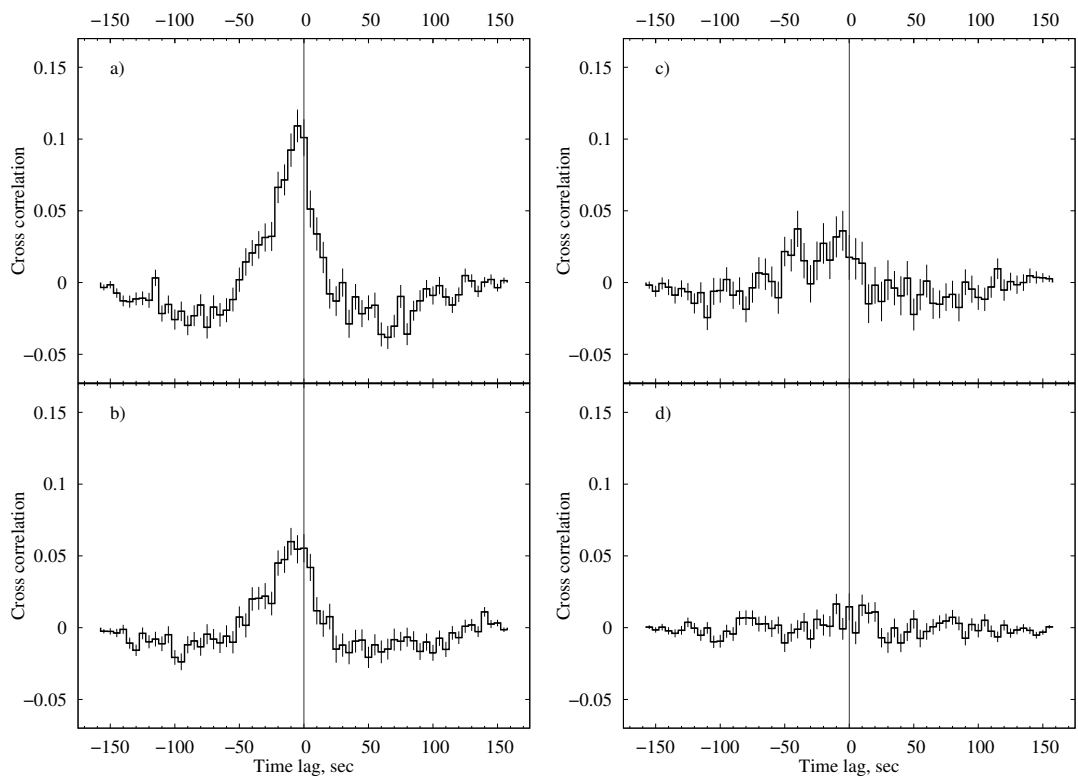}
\caption{Correlation functions of the X-ray light curves in the ranges of 2--5\,keV and 8--20\,keV for different precession phases (Table~\ref{tab:data_xray}): a~--- group I, b~--- II, c~--- III, d~--- IV. The CCFs were obtained by dividing the light curve into 150-s long intervals with subsequent averaging of individual CCFs. The shift of the CCF maximum in the negative direction indicates a lag of the soft emission behind hard emission.}
\label{fig:xray+xray}
\end{figure*}

\begin{figure*}
  \includegraphics[width=0.81\textwidth]{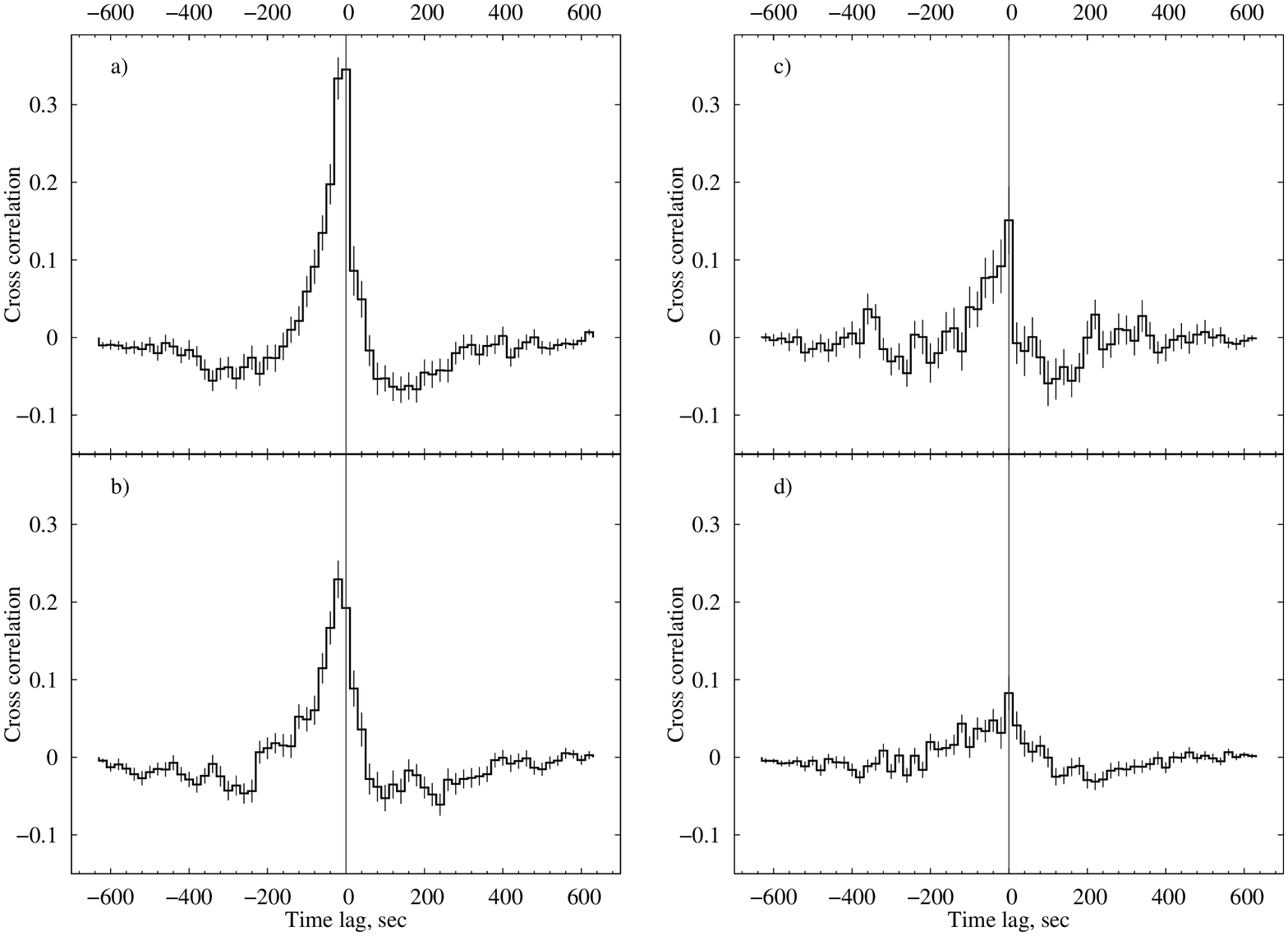}
\caption{Same as in Fig.~\ref{fig:xray+xray}, but for 640 s intervals.} 
\label{fig:xray+xray2}
\end{figure*}

\begin{figure*}
  \includegraphics[width=0.8\textwidth]{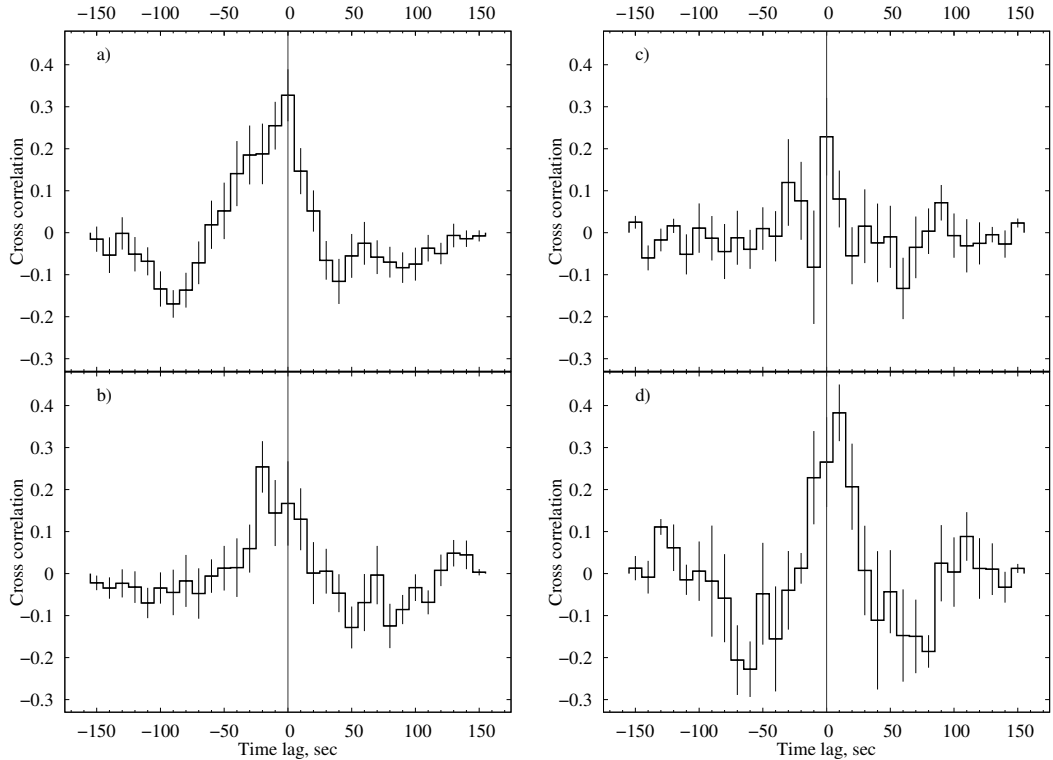}
\caption{Correlation functions of X-ray 2--20\,keV and optical light curves (Table~\ref{tab:data_xray+optics}). The shift of the CCF maximum in the negative direction indicates a lag of the (soft) X-ray emission with respect to the optical.} 
\label{fig:xray+optics}
\end{figure*}

\begin{figure*}
  \includegraphics[width=0.4\textwidth]{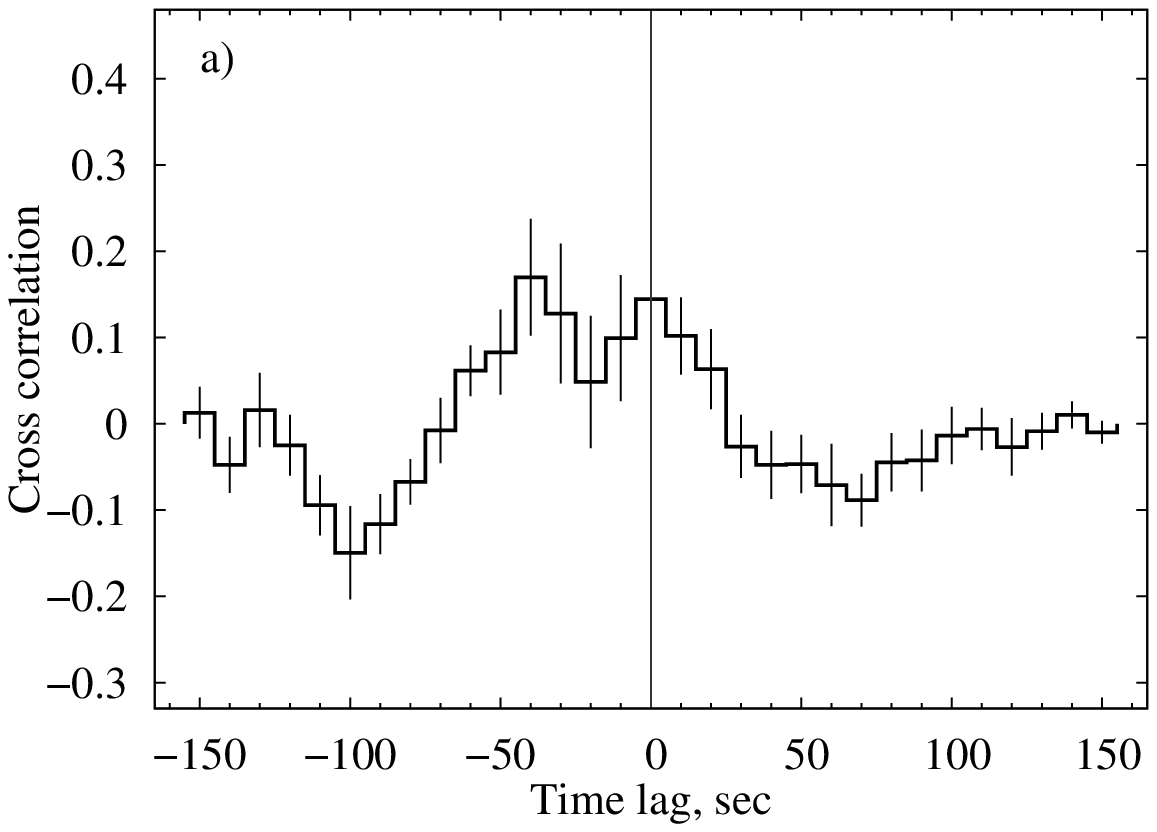}
  \includegraphics[width=0.4\textwidth]{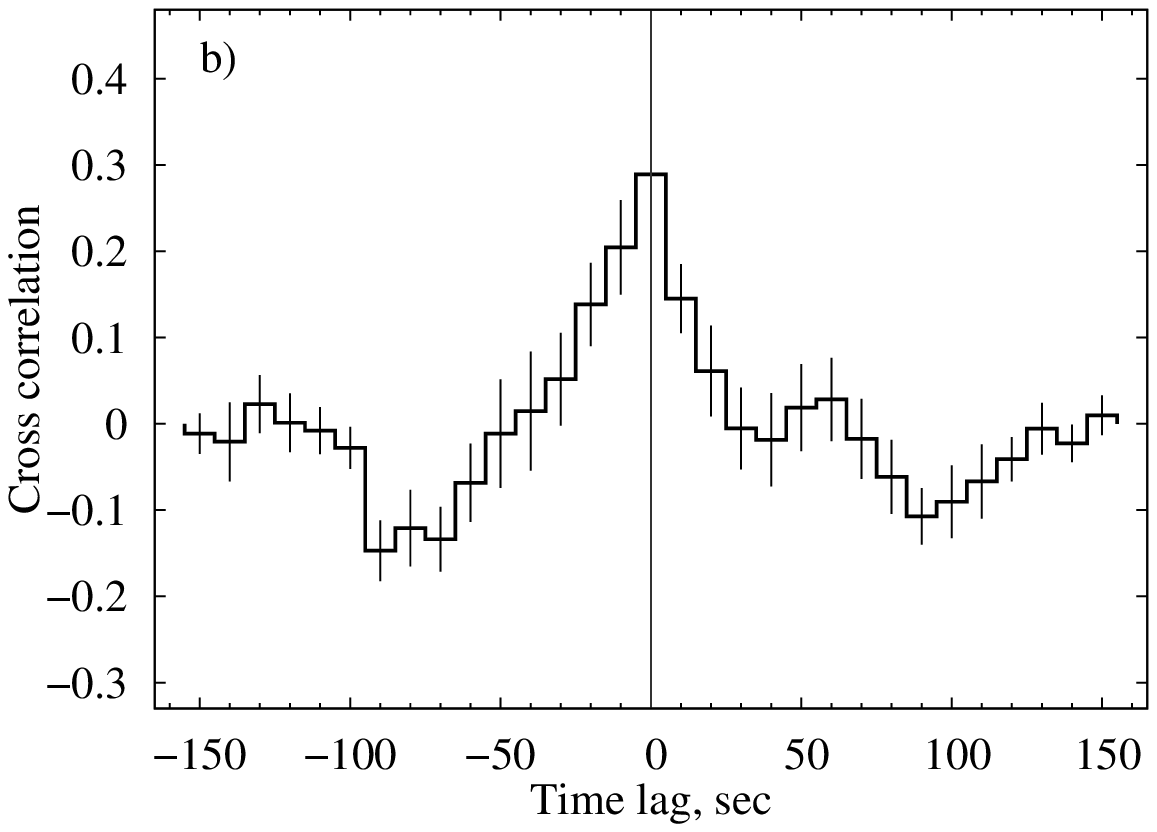}
\caption{Correlation between the X-rays at 2--5\,keV (a) and 8--20\,keV (b) and the optical based on the data of simultaneous RXTE and BTA observations, for 150-s intervals.} 
\label{fig:xray+bta}
\end{figure*}

\begin{figure*}
  \includegraphics[width=0.4\textwidth]{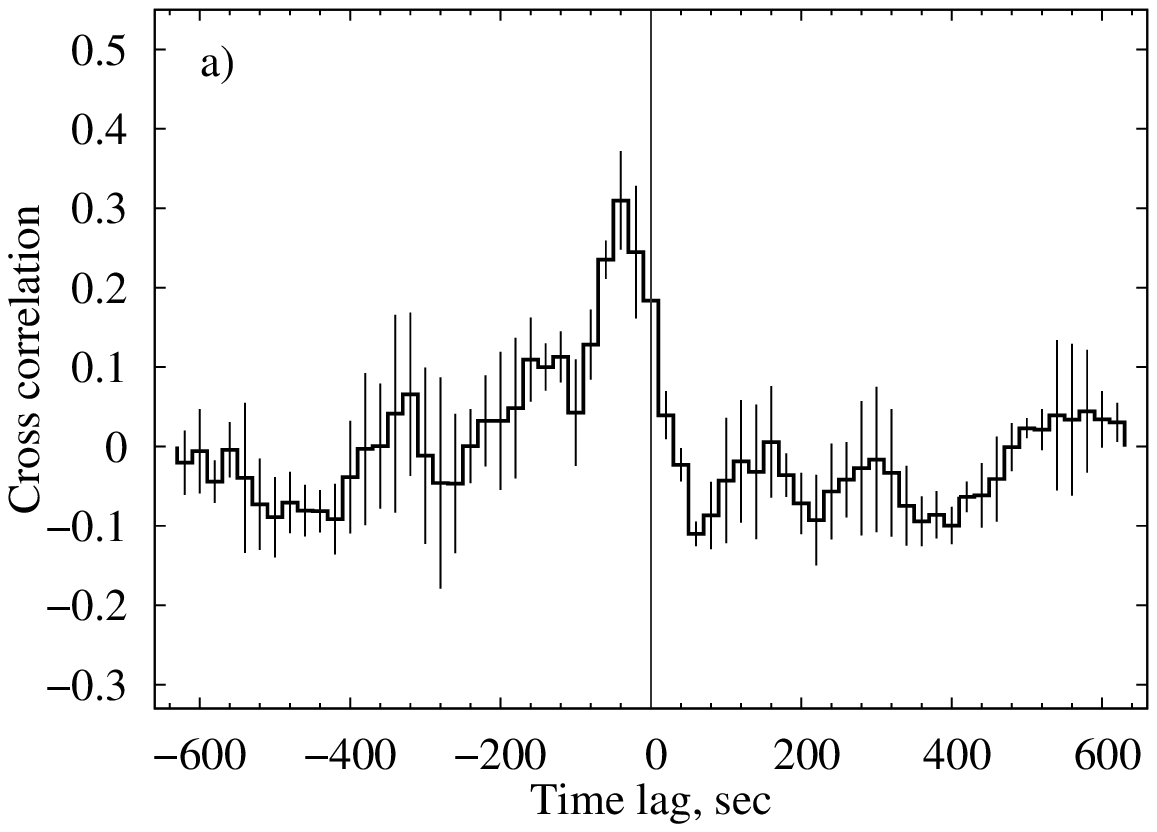}
  \includegraphics[width=0.4\textwidth]{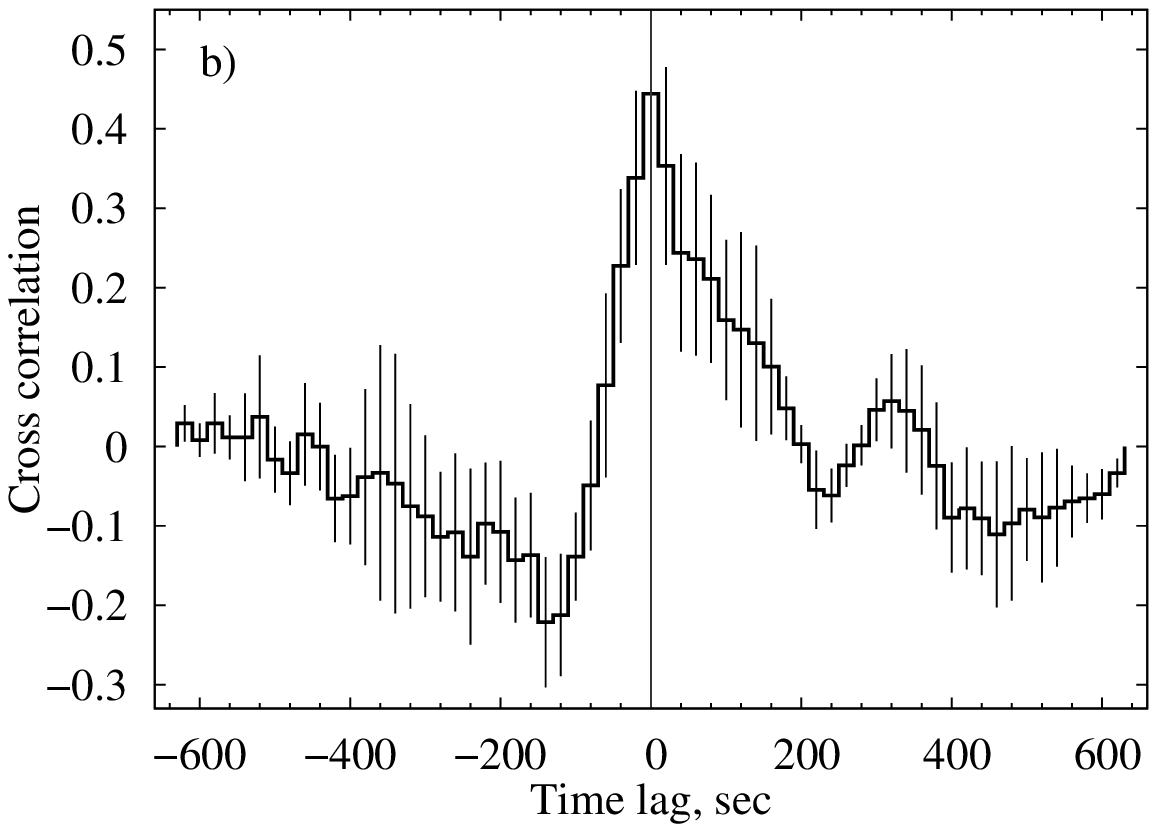}
\caption{Same as in Fig.~\ref{fig:xray+bta}, but for 640-s intervals.} 
\label{fig:xray+bta2}
\end{figure*}

We extracted the light curves in these two ranges, divided them into intervals of equal lengths, computed separate CCFs for each interval and then averaged. We found that the shape of the CCFs does depend on the interval length. In the case of shorter intervals, the structure of the CCF profiles is clearly seen, but tracing the correlation at large time-scales is not possible. On the other hand, increasing the length of the intervals causes new details corresponding to larger time-scales to appear on the CCF. That is why we thought it necessary to show in Fig.~\ref{fig:xray+xray} and \ref{fig:xray+xray2} the plots for 160~s and 640~s  interval length respectively.

Fig.~\ref{fig:xray+xray} shows cross-correlation between the X-ray light curves in the  2--5\,keV and 8--20\,keV ranges for 160~s intervals. The CCFs correspond to four precession-phase groups (Table~\ref{tab:data_xray}). The plots corresponding to the maximally open funnel show a pronounced peak at $-5$~s and a broad base spreading out to $-60$~s. Larger-scale trends are not visible with such interval length. The shift of the CCFs in the negative direction indicates a delay of the soft emission with respect to the hard. We see a similar pattern in the second-group (intermediate precession phases). The shape of the cross-correlation profile is unchanged, but the amplitude became smaller. The third-group demonstrates two separate peaks of approximately equal height at $-5$~s and $-40$~s. There is no statistically significant effect visible in the CCF of the fourth-group (edge-on disc).

The CCFs for 640~s intervals are shown in Fig.~\ref{fig:xray+xray2}. The resolution here is 20 s/bin, and the features at $-5$ and $-60$ s are not resolved well. However, it can be seen that the left wing of the profiles spreads out to 200--250 s. We found that the width of the profile stops to increase when the intervals become larger then $\sim640$~s. This indicates that 250~s is the maximum lag between the soft and hard emission, and that there is no correlation between them at larger time-scales. Gaussian analysis of the group~I CCF for these two intervals shows that there are three characteristic time-scales at $-5$, $-25$ and $-45$~s having the FWHM of 20, 50 and 150~s, respectively. 

The soft emission lags are possibly related to the fact that it originates mainly in the jets moving with the velocity of $v_j\simeq0.26c$, whereas the hard component is the emission reflected by the funnel wall. The idea that a lag should exist between the jet component and the reflected emission was first proposed by \cite{Revn2004}.

We assume that both the emission coming out from the funnel and the jet activity are determined by the processes in the proximity of the black hole. The orientation of SS\,433 is such that the observer does not see the inner parts of the funnel, and the jet can be seen only starting from a certain minimal distance $r_{j0}$. That is why the variability of the jets (and soft X-rays) will lag behind the variability of hard emission by a typical time $\tau\propto r_{j0}/v_j-r_{j0}\eta/c$, where $\eta$ is a coefficient depending on the geometry of the funnel.

The correlation functions of the X-ray 2--20\,keV and optical are shown in Fig.~\ref{fig:xray+optics}. The CCF maximum is located around zero. This indicates that both the X-ray and the optical emissions of SS\,433 form in the same place, specifically, in the funnel of the supercritical accretion disc. Figs.~\ref{fig:xray+optics}b,c,d correspond to the observations when the funnel is best visible. But these observations, unfortunately, are very noisy. The asymmetry of the CCF which is clearly seen in the BTA data (Fig.~\ref{fig:xray+optics}a) is already visible in the comparatively long observation from the 1.5-m telescope (Fig.~\ref{fig:xray+optics}b).

Fig.~\ref{fig:xray+optics}a represents the third group and corresponds to the longest observation. For this data set we may separately investigate the correlation between the optical emission and the soft and hard X-rays. The results are shown in Fig.~\ref{fig:xray+bta}a for 2--5\,keV (jet range) and Fig.~\ref{fig:xray+bta}b for 8--20\,keV (funnel range). The CCF profile of the hard X-rays and optical is symmetric and has a peak strictly at zero, i.e., hard X-ray emission is synchronous with the optical. The correlation between soft X-rays and optical has a bimodal profile and is very similar in shape to the correlation function between the soft and hard X-rays for the same precession phase (Fig.~\ref{fig:xray+xray}c).

Such a behaviour is consistent with the idea of \cite{Revn2004} that the optical emission is generated due to the thermal reprocessing on the funnel walls. Apparently, the hard emission formed in deepest regions of the funnel is partly reflected and partly reprocessed into the optical range in the outer regions of the funnel. So it could be possible that the hard X-ray and the optical emissions are formed in the same place and therefore reach the observer about simultaneously. That is why the correlation functions between the soft and hard X-rays and between soft X-rays and the optical appear practically identical (Fig.~\ref{fig:xray+xray}c and ~\ref{fig:xray+bta}a).

Fig.~\ref{fig:xray+bta2} shows the correlations between the two X-ray components and the optical for 640~s intervals. Here we again see the lag of the soft X-ray emission behind the optical. The CCF for the soft X-rays and the optical (Fig.~\ref{fig:xray+bta2}a) again demonstrates a great similarity to the CCF of the soft and hard X-rays (Fig.~\ref{fig:xray+xray2}c). However the correlation between the hard X-rays and the optical appears different (Fig.~\ref{fig:xray+bta2}b). The profile has a clear asymmetry to the right, and a side peak at 300 sec. This may indicate that a certain portion of hard X-ray emission is ahead of the optical, or a part of the optical emission lags behind the X-rays. The 300 sec delay of optical emission might be related to the heating effects of the funnel wall by the hard emission which appears at these time-scales.

Thus, the delay of soft X-rays may be interpreted as a lag of the jet emission relative to the reflected/reprocessed emission (hard X-rays and optical). Within the framework of our simple geometric model of the funnel we can estimate this delay. Both the visible jet base $r_{j0}$ and the geometric coefficient $\eta$, which accounts for the scattering of emission in the funnel, should vary with precession phase. The position of the CCF peak must change from $-20$~sec for the phase when the funnel is best visible to $-80$~sec for close-to-`edge-on' phases. 

A similar effect is observed. We see a lag of up to 100~s of soft X-rays behind the optical on the correlation functions for the optical and X-ray emission (Fig.~\ref{fig:xray+optics}--\ref{fig:xray+bta2}). This effect can also be seen in the CCFs of soft and hard X-rays (Fig.~6, 7). A Gauss analysis of the CCF profile corresponding to the maximal opening of the disc (Fig.~\ref{fig:xray+xray}a, \ref{fig:xray+xray2}a) showed that one of the components is located at $-25$~s. A feature at $-(50-100)$~s can be seen at other precession phases (Fig.~\ref{fig:xray+xray}b,c). However, the structure of the CCF cannot be fully explained within the framework of this simple geometric model. Moreover, it is impossible to explain the lag of the soft emission at times up to $\sim200$~s in Fig.~\ref{fig:xray+xray2}.

\subsection{Jet model}
\label{sec:jet_model}

\begin{figure*}
  \includegraphics[width=0.4\textwidth]{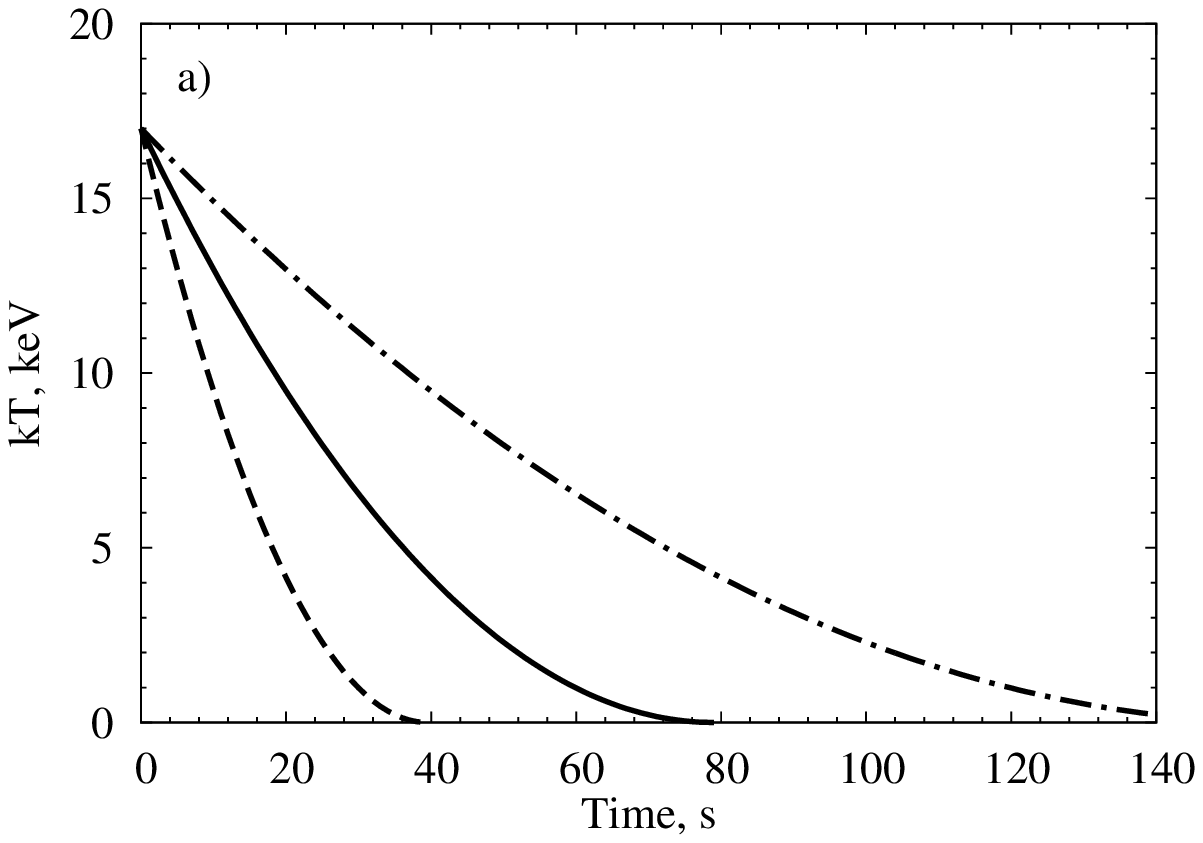}
  \includegraphics[width=0.4\textwidth]{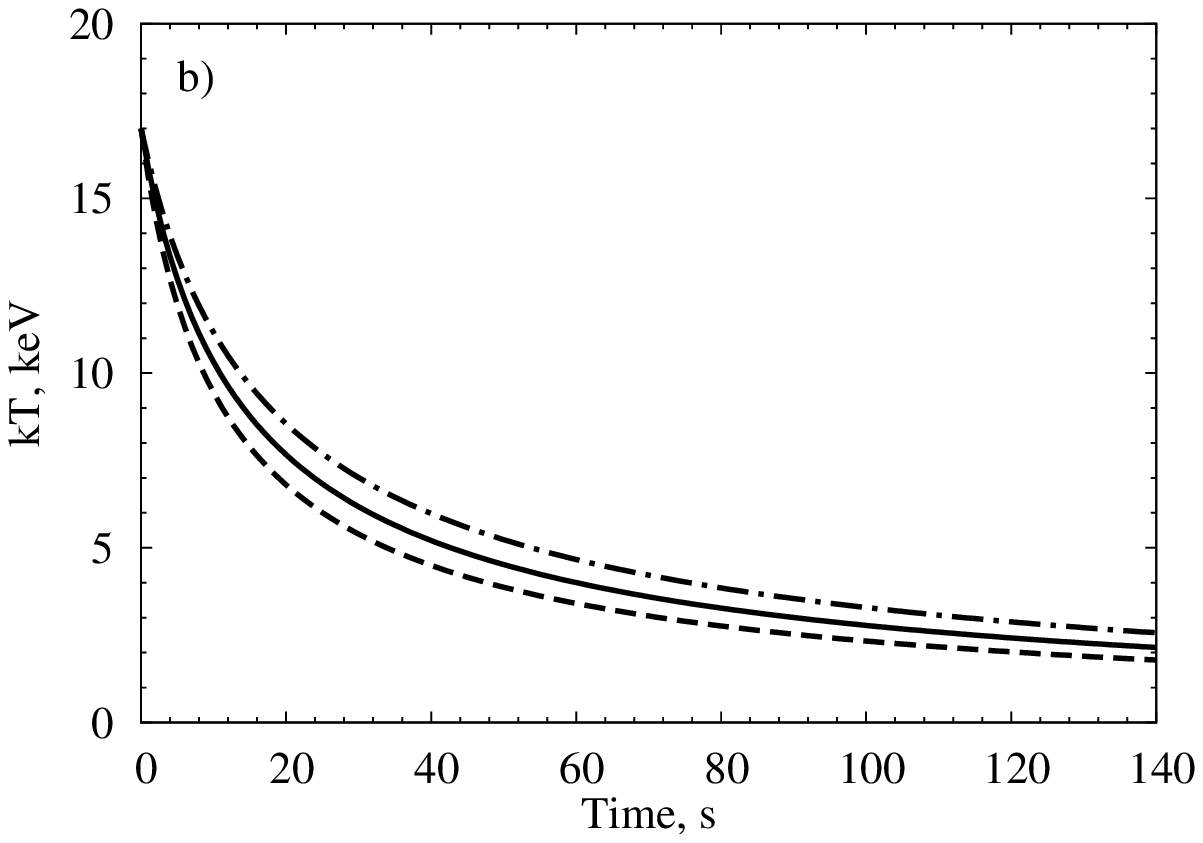}
\caption{Dependence of the jet blob temperature on time for the cases of radiative (a) and adiabatic (b) cooling. Different lines show different initial densities: solid~--- $4\times10^{13}$ cm$^{-3}$, dashed and dotted~--- $2\times10^{13}$ cm$^{-3}$, and dashed~---  $8\times10^{13}$ cm$^{-3}$. The initial temperature is $\theta_0=17$~keV.}
\label{fig:jetmodel_temp}
\end{figure*}

\begin{figure*}
  \includegraphics[width=0.4\textwidth]{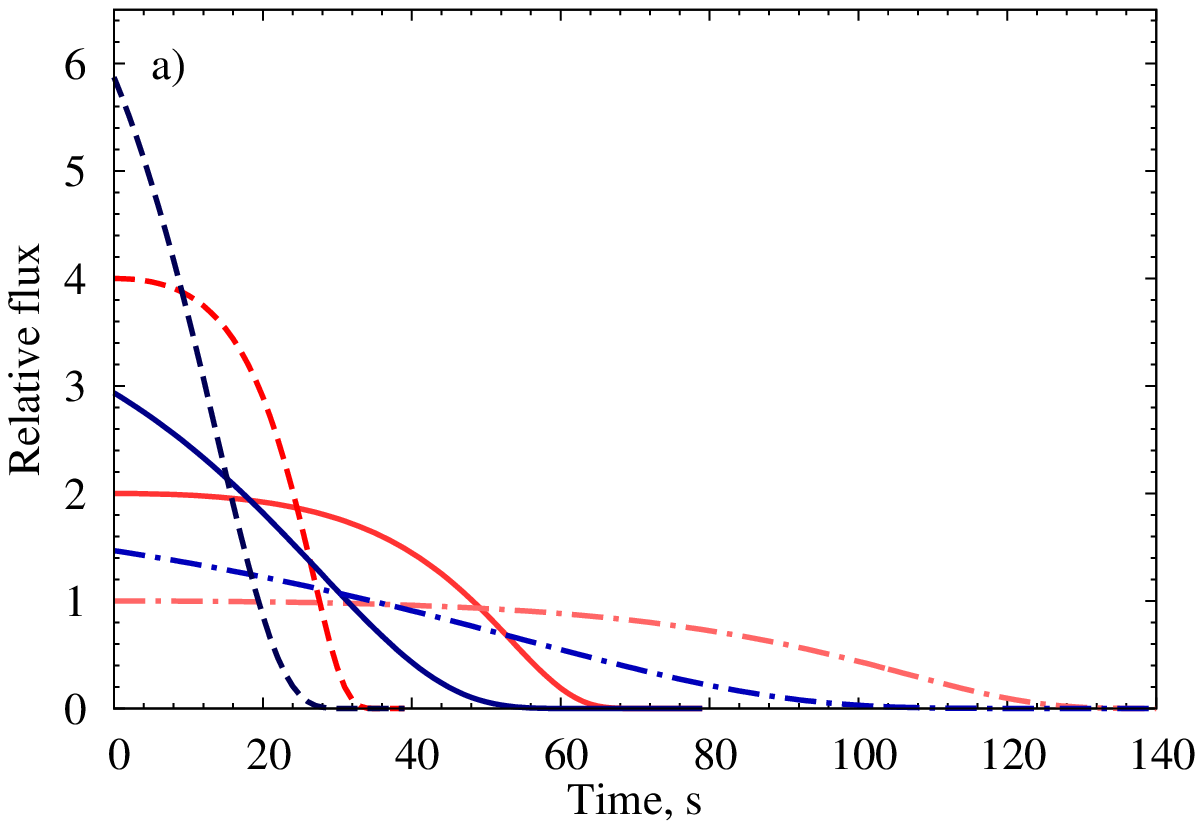}
  \includegraphics[width=0.4\textwidth]{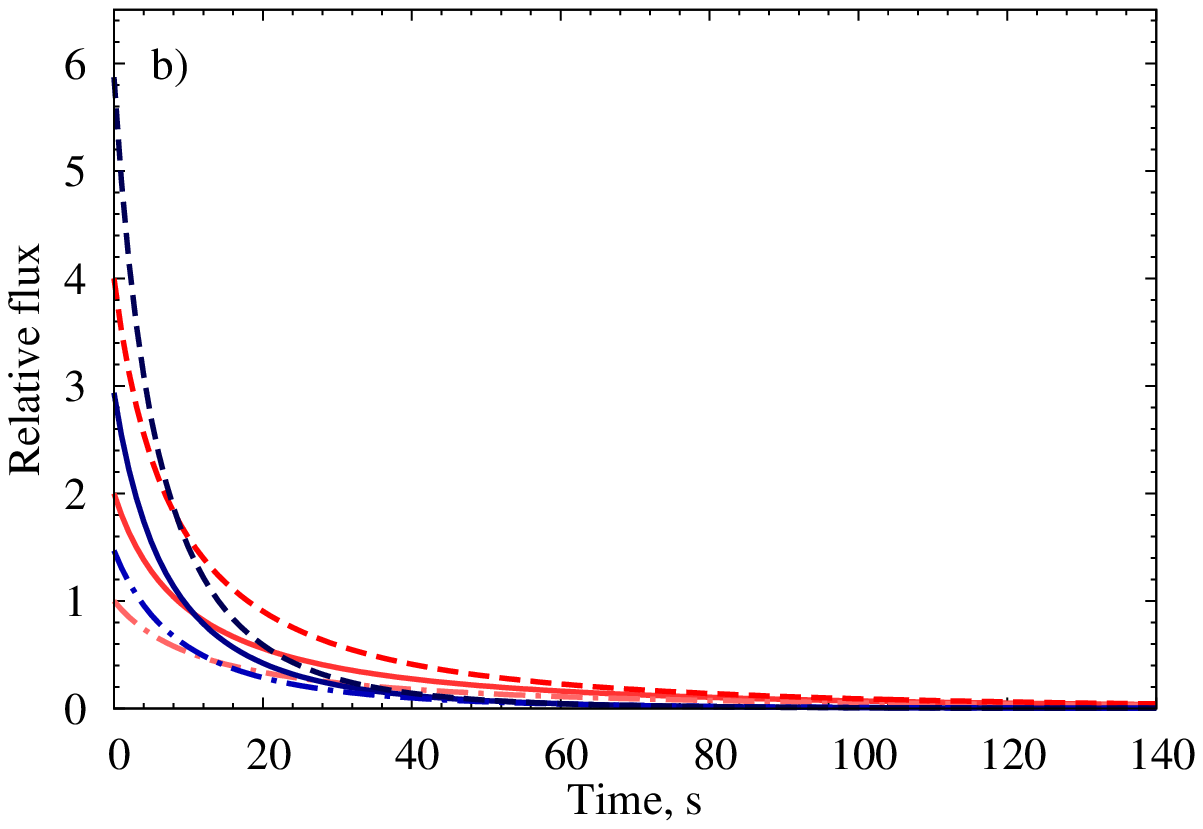}
  \caption{Dependence of the X-ray emission flux of the blob on time for the cases of radiative (a) and adiabatic (b) cooling. The flux in the 2--5~keV range in shown in grey (red), and the flux in the 8--20~keV range is shown in dark grey (blue). Different initial densities are shown by various types of lines, as in Fig.~\ref{fig:jetmodel_temp}. The 2--5~keV flux of a blob with a density of $2\times 10^{13}$~cm$^{-3}$ at $t=0$ is taken as unity. (A colour version of this figure is available in the online journal.)}  
\label{fig:jetmodel_flux}
\end{figure*}

In this section we discuss an additional (or alternative) explanation of the CCFs which is based on the cooling of the relativistic jets. Spectral studies of the jets show that the temperature at the visible jet base is $k_BT_0\approx17$~keV \citep{MedvFabr2010}. The gas of the jets cools due to expansion and radiative losses. The 2--5~keV range contains most of the jet emission, but a certain portion should also be emitted in the 8--20~keV range. In this case, both in the hard and soft range, we should see the emission of the same clouds (blobs), which make up the jets. A jet moving towards us is much brighter than the one moving away and contributes more to the total flux; in addition, the receding jet may be partially obscured from the observer by the wind \citep{MedvFabr2010}. Therefore, the light curves in the two ranges should be similar and well correlated with each other. The CCF asymmetry in this case is related to the fact that as the jet cools, the hard-range flux should weaken faster than in the 
soft range. Having modeled the cooling of the jet gas, one can try to reproduce the shape of the CCFs.

In our model we assumed that the jet consists of spherical blobs, which are shot out from the inner parts of the accretion disc with a given time-scale. The blobs are optically thin, and are dominated by bremsstrahlung emission. The volume emission coefficient may be written in the following form \citep[Sect. 5.2]{Rybicki} (erg cm$^{-3}$ s$^{-1}$ Hz$^{-1}$): 
\begin{equation}
\varepsilon_\nu = 6.8\times 10^{-38} Z^2 n_e n_i T^{-1/2}e^{-hv/kT} \langle g_{ff} \rangle  
\end{equation}
We now introduce $\theta = k_B T$ and $\epsilon = h\nu$. The Gaunt factor $\langle g_{ff} \rangle$ is roughly described by the expression $(\epsilon/\theta)^{-0.3}$ \citep{Culhane1970} in the X-ray range. We can introduce the plasma emissivity, $J_\epsilon = \varepsilon_\epsilon/n^2$, which does not depend on density:
\begin{equation}
J_\epsilon (\theta) \propto \frac{1}{\sqrt{\theta}}\left(\frac{\epsilon}{\theta}\right)^{-0.3} e^{-\epsilon/\theta} \mbox{erg cm$^{3}$ s$^{-1}$ keV$^{-1}$}
\label{eq:emis}
\end{equation}

In the general case, the temperature $\theta(t)$, density $n(t)$, and blob size $r_b(t)$ are functions of time. The flux from one blob in the energy range from $\epsilon_1$ to $\epsilon_2$ is expressed by the formula:
\begin{equation}
F_{\epsilon_1,\epsilon_2}(t) = \frac{4\pi}{3} n(t)^2 r_b(t)^3\int\limits_{\epsilon_1}^{\epsilon_2} J_\epsilon(\theta(t))d\epsilon
\end{equation}

For the sake of simplicity we assumed that all the blobs in the jet have the same initial size and density. In this case, the initial size $r_{b0}$ and density $n_{0}$ are related through the kinetic luminosity of the jets:
\begin{equation}
L_k = \frac12 M_b \langle\dot{N}\rangle v_j^2 = \frac{2\pi}{3} \mu m_p \langle\dot{N}\rangle n_{0} r_{b0}^3 v_j^2,
\label{eq:blobsize}
\end{equation}
$\langle\dot{N}\rangle$~--- the average number of blobs per second, $M_b$~--- mass of a blob, kinetic luminosity $L_k\sim10^{39}$~erg/s \citep{Panferov1997}. In all computations we assumed that $\dot{N}$ obeys a Poisson distribution with an average of 1 blob per second.

Depending on the conditions in the jet, different cooling mechanisms may be realized. The blobs may be enveloped in external gas. They therefore could have approximately constant sizes, because the pressure of the external medium would prevent their expansion. In this case, the cooling occurs due to radiative losses. If the external medium is absent or its pressure is not enough to constrain the expansion, the cooling is possible both due to radiative losses and expansion. Depending on the density of the blob, one or other mechanism prevails. We therefore considered two extreme cases: (i) non-expanding blobs, cooling due to radiative losses; 
(ii) blobs expanding with the speed of sound and cooling adiabatically. For a detailed review of the behavior of the temperature and other blob parameters for these two models see the Appendix.

Fig.~\ref{fig:jetmodel_temp} shows the time dependences of temperature for these two models. The blobs with higher initial densities cool faster in both models. However, the dependence of the cooling rate on density is stronger in the radiative model. We can introduce the typical cooling times $t_e^{rad}$ (\ref{eq:te_rad}) and $t_e^{ad}$ (\ref{eq:te_ad}). They are e-fold times which depend on the initial temperature and density of the blob. We found that at a density of $n_{cr}\sim4\times10^{13}$ cm$^{-3}$, the cooling times of two models become comparable and equal to $\approx30$~s. Both mechanisms work equally efficiently with this initial density. At a density of $n \gg n_{cr}$, the radiative cooling mechanism dominates, whereas the adiabatic mechanism prevails at $n \ll n_{cr}$.

In Fig.~\ref{fig:jetmodel_flux} we show the dependence of the X-ray flux in the 2--5~keV and 8--20~keV ranges on time for different initial densities. The flux shows a steeper decline in the hard range compared to the soft range, both in the radiative and adiabatic cases. Moreover, in the radiative model, the flux profiles in the hard and soft ranges differ significantly from each other. While the temperature is sufficiently hot, the soft range flux remains practically unchanged. In the hard range, it begins dropping straight away. The adiabatic model shows no such sharp difference of profiles, since the weakening of the flux is related not only to the decrease in temperature, but also to the decrease of the blob density.

\begin{figure}
  \includegraphics[width=0.49\textwidth]{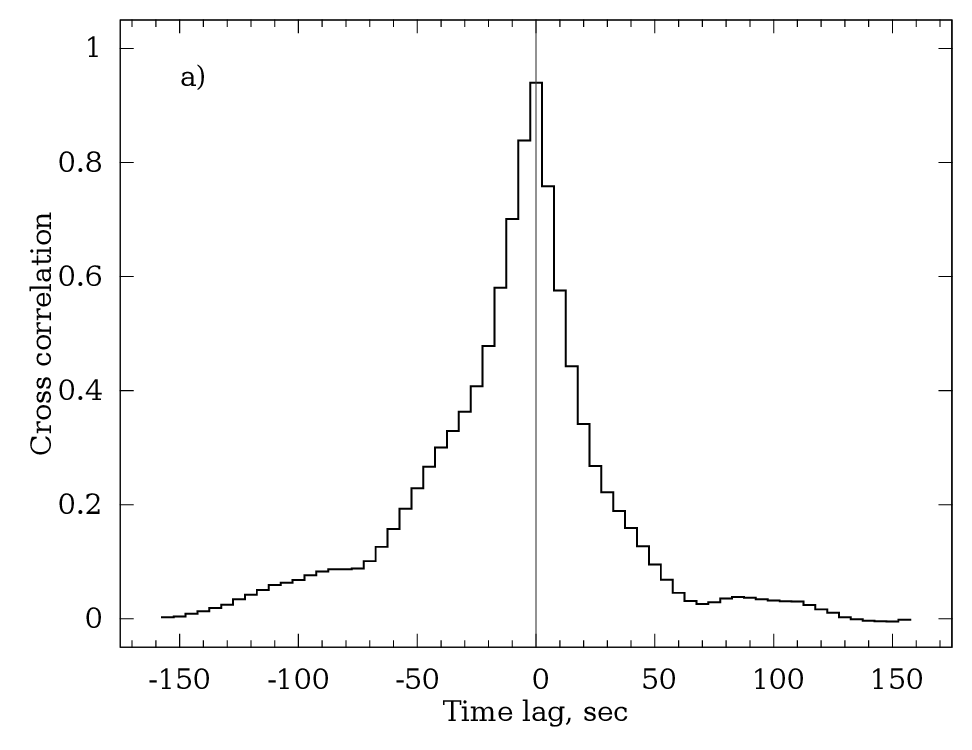}
  \includegraphics[width=0.49\textwidth]{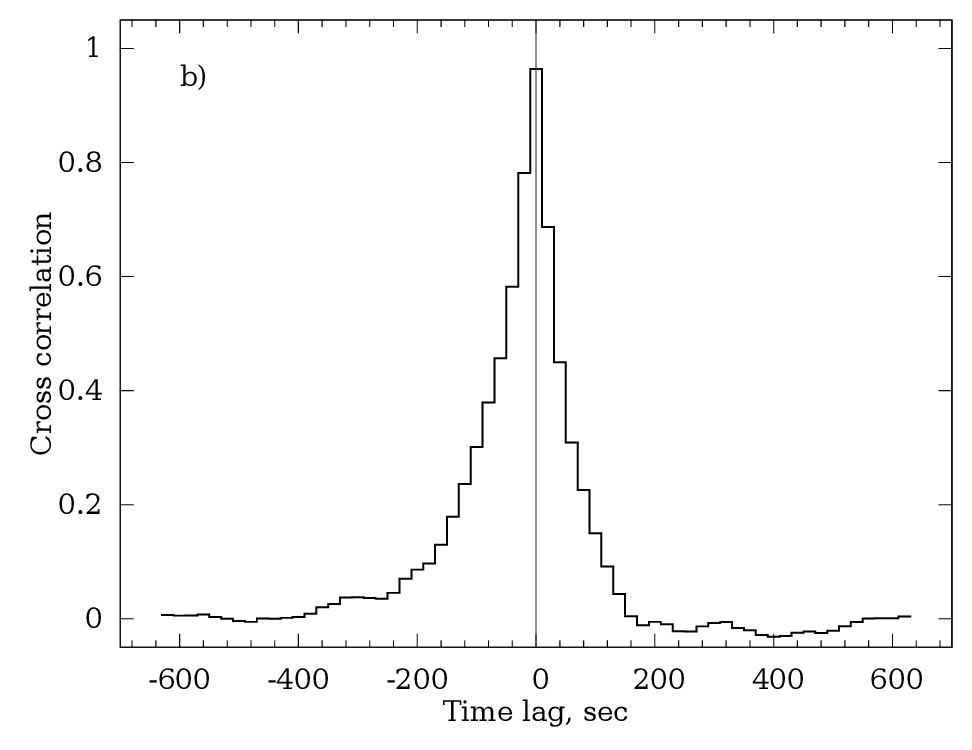}
  \includegraphics[width=0.49\textwidth]{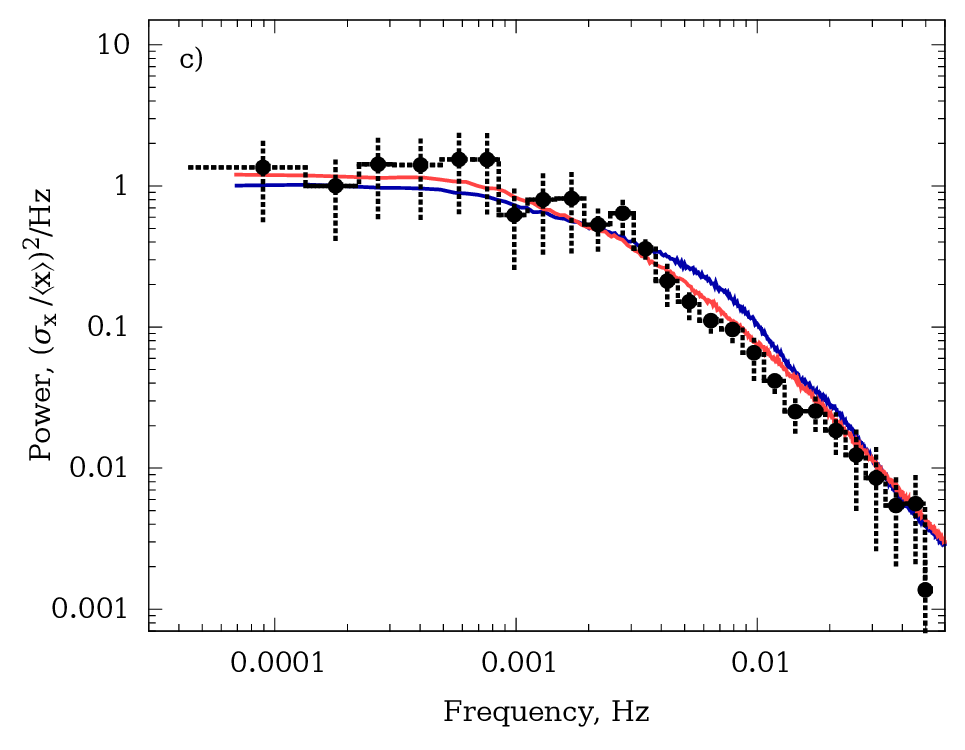}
\caption{Cross-correlation functions of 2--5 and 8--20~keV light curves, obtained in the cooling jet model for 150~s (a) and 640~s (b) interval length. (c)~--- the observed (circles) and model ( dark grey/blue line) power spectra in the 2--20~keV range. In these cases, three-component model was used, consisting of three blob fractions, but the grey (red) line represents the power spectrum of one-component model (the lowest density fraction, see text). (A colour version of this figure is available in the online journal.)}  
\label{fig:jet_results}
\end{figure}
 
Synthetic light curves were computed as follows. We assumed that all the blobs in the jet are the same. We used a random number generator to set the number $\dot{N}$ of ejected blobs, followed the flux evolution of each blob and added them into the light curve. Further, we computed the model CCFs using synthetic light curves in two energy ranges. 

We were able to reproduce the asymmetry of the observed CCFs. The asymmetry is more conspicuous in the radiative cooling model than in the adiabatic model. This is due, mainly, to the difference between the flux decrease rates in the soft and hard ranges in the radiative model mentioned above. The width of the CCF peak in each model is determined by the corresponding typical cooling time: $t_e^{rad}$ or  $t_e^{ad}$.

We found above that the observed CCFs of SS\,433 have a complex profile (Fig.~\ref{fig:xray+xray} and Fig.~\ref{fig:xray+xray2}). When the disc is maximally open to the observer, we can speak of at least three components with the FWHM of 20, 50 and 150 s. It is impossible to fully reproduce the shape of the observed CCF using blobs of only one type. Each density value corresponds to one typical cooling time and can describe only one CCF component. We therefore considered a combined model, where the jet consists of three groups of blobs (hereafter, we refer to them as `fractions' A, B, and C) with different initial densities corresponding to three different typical times.

Blobs from different fractions may be ejected with different frequencies $\langle\dot{N}\rangle$ and may contain different portions of the jet kinetic luminosity $L_k$. Here we adopted that the blobs of each fraction are ejected independently, $\langle\dot{N}\rangle=1$ for each fraction, and also that equation (\ref{eq:blobsize}) remains true for all fractions. These assumptions allow us to derive the combined model light curve by means of coadding with the weights of the light curves corresponding to individual fractions. The weights are needed to account for the amplitude differences of the three observed CCF components. Actually one can restore the distribution of the blobs by fractions, however it was not our aim. 

Fig.~\ref{fig:jet_results}a,b shows the model CCFs. They reproduce the shape of the observed CCFs of SS\,433 corresponding to the precession phases of the open disc (Figs.~\ref{fig:xray+xray}a and \ref{fig:xray+xray2}a) fairly well. We used a jet model consisting of three fractions of blobs with initial densities at $r_{0j}$: $8\times 10^{13}$~cm$^{-3}$ (fraction A), $3\times10^{13}$~cm$^{-3}$ (B) and $5\times10^{11}$~cm$^{-3}$ (C). The fraction A with the highest density is responsible for the formation of the narrowest and highest CCF component. We used the radiative cooling model for this fraction. Fraction B with the density $3\times10^{13}$~cm$^{-3}$ is responsible for the formation of the middle component. For this fraction we also used the radiative model, in spite of the fact that its initial density is close to critical. The radiative model reproduces the asymmetry of the profile better. The lowest-density fraction C well describes the broadest CCF component. For this fraction, we used the adiabatic 
model, since its initial density is $5\times10^{11}$~cm$^{-3}$~$\ll n_{cr}$. To reconstruct the relation between the amplitudes of the CCF peaks, we had to assign the highest weight to the lowest-density fraction C. Thus, within the framework of the cooling jet model, we can say that most of the jet mass falls within this fraction.

The cooling jet model allows constructing the power spectra. Recall that the observed PDS of SS\,433 exhibit a break at the frequency $1.7\times 10^{-3}$~Hz (Fig.~\ref{fig:pds_data}). The observed PDS is practically flat in the low-frequency range, whereas in the high-frequency range it is described by a power law with the exponent 1.67.

Our stochastic jet model is a type of shot noise, where individual shot profiles (Fig.~\ref{fig:jetmodel_flux}) are determined by the blob parameters and the cooling time.
Both the adiabatic and the radiative models reproduce the break in the power spectrum. The frequency of the break is determined by the typical cooling time, i.e., it depends on the initial temperature and density. The slope of model PDS at high frequencies does not depend on the initial conditions in the jet and is determined solely by the cooling mechanism. The slope in the adiabatic model does represent slope of the observed PDS, 
whereas the radiative model gives a steeper PDS with the index $\approx 2$. 



In the $f \ll t_{e}^{-1}$ frequency range the shape of the power spectrum is no longer influenced by the specific cooling mechanism. The main role is now played by the dynamics of the jet formation process. We assumed above that the blob ejection frequency $\dot{N}$ obeys the Poisson distribution, and its mean value does not depend on time. That is why in our case both cooling models yield a flat power spectrum (white noise) at low frequencies. If one adopts that $\langle\dot{N}\rangle$ as well as initial density, blob temperature or amount of matter ejected in the jet vary with time, then the power spectrum at low frequencies may turn out to be different.

Fig.~\ref{fig:jet_results}c shows the power spectrum for the combined model, consisting of three fractions. It describes the observational data well. We used the model with the same densities that were used to describe the correlation functions (Fig.~\ref{fig:jet_results}a,b). The shape of the model PDS mainly determined by the contribution from the lowest-density fraction~C with the density of $5\times10^{11}$~cm$^{-3}$. The power at $0.004-0.02$~Hz is slightly overestimated due to fraction B with a density of $3\times10^{13}$~cm$^{-3}$. The most dense fraction A does not contribute to the observed frequency range ($f<0.05$~Hz). By varying the density of fraction B and the contribution of this fraction to the total jet mass, one can achieve a better agreement between the observed and theoretical power spectra. However if we take only one fraction~C (grey line in Fig.~\ref{fig:jet_results}c) we already obtain a very good agreement between model and observed PDS. Thus we conclude that the shape of the power 
spectrum is determined mainly by the fraction C with a minor contribution by fraction B.  

Thus, the cooling jet model well explains the rapid X-ray variability of SS\,433. To adequately describe the main typical features of the CCFs and PDS, we required three fractions of blobs with different densities. \citet{Marshall2013} basing on the Chandra observations showed that the density of the jets should be in the range of $10^{10}-10^{13}$~cm$^{-3}$. The densities obtained by us agree fairly well with their estimates. Further refinement of the model will allow to impose stricter limits on the jet density and describe better the observed data.

\section{Discussion}
\subsection{Model comparison: funnel vs jet}
Above we described two mechanisms that may explain the nature of rapid variability of SS\,433. The first considers the reflection and scattering of emission in the wind funnel of the supercritical accretion disc (hereafter we refer to it as funnel model). The second~--- the process of cooling of the gas that composes the jets (jet model). Each of the models may explain most of the observed facts. 

The funnel model (Sec.~\ref{sec:funnel_model}) well reproduces the position of the break and the slope at high frequencies in the power spectrum corresponding to precession phases of the open disc (Fig.~\ref{fig:pds_results}c). The frequency of the break is determined by the length, and the slope of the power spectrum at high frequencies~--- by the opening angle of the funnel. However the funnel length ($l_f \sim 5\times 10^{12}$~cm) and the jet base radius ($r_{j0} \sim 6\times 10^{11}$~cm) derived by us are about $1.5-2$ times larger than the corresponding values obtained in other studies \citep{MedvFabr2010,Cherepashchuk2005}.

The funnel model is capable to explain the lag of soft X-ray emission behind the hard X-ray and optical emission. It is assumed that most of the soft X-rays come from the jets. The observed hard X-rays and optical are reflected/reprocessed by the funnel wall. The duration of the delay should be determined by the jet velocity, the funnel geometry and its orientation in space (precession phase). The model predicts a 20--80~sec delay depending on the precession phase, but cannot fully reproduce the shape of the CCFs.

The jet model (Sec.~\ref{sec:jet_model}) well reproduces both the X-ray CCFs and the PDS (Fig.~\ref{fig:jet_results}). In the jet model, the observed features of the correlation functions and power spectra are related to the typical cooling times of the clouds that make up the jet. The higher the cloud density, the shorter the cooling time. We considered the model of a jet consisting of three fractions with different densities corresponding to the CCF components. The two most dense fractions are responsible for the formation of two narrow CCF components, whereas the rarefied fraction~--- for the broad component and the power spectrum.  

The jet model can explain even the amplitude difference between the hard and soft X-ray PDS (Fig.~\ref{fig:pds_data_g1g4}a). As the jets cool, the hard X-ray flux changes more abruptly~(Fig. \ref{fig:jetmodel_flux}b), and therefore the power spectrum of the variability in the hard range should be approximately 2.5 times higher than in the soft range. Precisely this effect is observed in Fig.~\ref{fig:pds_data_g1g4}a.

Thus, the funnel model explains the break in the X-ray power spectrum and its slope at the frequencies above the break but it overestimates the values of $l_f$ and $r_{j0}$. It explains qualitatively the CCFs in the X-ray (soft~-- hard) and optical ranges. On the other hand, the jet model can also explain the break in the X-ray PDS. Moreover, the model yields expected values of gas density in the jet clouds. In addition, this model well describes the shape of the X-ray CCF and explains the amplitude difference between the soft and hard range PDS. However, the jet model cannot explain the optical variability.
 
The jets have a temperature of 17~keV at their base, and their emission should fall within the X-ray range. The X-ray luminosity of SS\,433 is $\sim10^{36}$~erg/s \citep{MedvFabr2010}. The V-band optical luminosity is two orders of magnitude higher, $\sim10^{38}$~erg/s \citep{Fabrika2004}. The contribution of the X-ray jets to the optical luminosity must be negligibly small. The optical jet luminosity is comparable to X-ray one but it is emitted in the lines of hydrogen \citep{Panferov1997}, and this emission is generated at considerably greater distances from the system ($\sim 10^{15}$~cm). The jet optical lines are appeared due to the dynamic interaction of the jet clouds with the gas of the surrounding wind.

The optical PDS \citep{Burenin2011} are very similar to those in X-rays obtained by us, they also exhibit a break. This task is well handled by the funnel model \citep{Revn2004}. The parameters of the optical PDS found by \cite{Burenin2011} and averaged over all observations (precession phases) are the following: the power-law index at low frequencies $\beta_1=1.15$, at high frequencies $\beta_1+\beta_2=2.95$, the break frequency $f_{br}=2.43\times10^{-3}$~Hz (equation~\ref{eq:pds_knee}). Our modelling of this PDS from \cite{Burenin2011} yields the funnel wall length $l_f\approx4.5\times10^{12}$~cm and opening angle $\vartheta_f\approx52\degr$. Optical power spectra are steeper than those in X-rays, and therefore the accuracy of the determination of angle $\vartheta_f$ is lower, $\sim 10^\circ$. 

We found that the power spectra of individual optical observations can differ significantly from the average. In particular, we constructed the power spectra for two longest data sets (2005-07-19 and 2005-07-23) from \cite{Burenin2011}. Their parameters are: $\beta_1=1.02$ and $1.29$, $\beta_1+\beta_2=3.07$ and $2.61$, $f_{br}=3.8\times10^{-3}$~Hz and $4.3\times10^{-3}$~Hz, respectively. The break frequency for these two individual PDS is 1.5 times higher than the mean value. Accordingly, individual observations yield smaller funnel sizes. We may estimate that in the jet direction the funnel size is $l_f\cos\vartheta_f\sim (2-2.5)\times 10^{12}$~cm which is comparable with donor star and the thick disc size \citep{Fabrika2004,Cherepashchuk2005}.

We can conclude from the above that both mechanisms contribute to the observed fast variability of SS\,433. Whereas in the optical range variability must be determined only by the funnel, in the X-rays we see both the manifestation of the funnel and the jets. The X-ray light curves are a sum of two varying signals: the light curve of the jets and the light curve of the funnel. Moreover, each of these signals has its own nature of variability and power spectrum. The smearing out of variability in the funnel is determined by its geometry and has to be the same in X-rays and optical. We can therefore expect that in the X-rays the intrinsic power spectrum of the funnel is a roughly similar to the observed optical PDS.

It follows from a comparison of the X-ray and optical power spectra (Fig.~\ref{fig:pds_data} here and Fig.6 in \cite{Burenin2011}) that the variability amplitude in the optical is lower than that in the X-rays. At the frequencies $>10^{-3}$~Hz, the amplitude in the optical is lower by one order of magnitude. This is probably due to the fact that the funnel itself is a less variable source than the X-ray jets. We conclude that the observed X-ray PDS is, for the most part, the PDS of the jets, since the funnel should contribute considerably less to the X-ray variability. Nonetheless, we have seen that the contribution from the funnel shows up in the CCFs between the X-rays and optical (Figs.~\ref{fig:xray+optics},\ref{fig:xray+bta}). Thus, the observed variability of SS\,433 can be explained only by the combined effect of two mechanisms: the cooling of the X-ray jets and the smearing out of variability in the inner parts of the funnel in the wind of the supercritical accretion disc. 

\subsection{Power spectrum at the `edge-on' precession phase}
In the precession phases when the disc is observed 'edge-on', we see a fundamentally different picture. The PDS (Fig.~\ref{fig:pds_data}) exhibits no break and can be described by a single power-law with an index of $1.34$ at all frequencies. Moreover, this PDS are the same both in the hard and soft ranges (Fig.~\ref{fig:pds_data_g1g4}b).

The hot inner parts of the funnel and jets are obscured for the observer by the funnel walls (the wind) in this precession phase. Given the size of the funnel from optical data $l_f\approx 4.5\times 10^{12}$~cm, as it was found in previous section, we estimate $r_{j0}\sim 7\times 10^{11}$~cm (visible jet base radius at the open disc phase). It corresponds to $\sim 250$~s of extra jet propagation time to be visible to the observer at the `edge-on' phase. In this time, even the least dense fraction manages to cool down to $\approx 4$~keV. At such temperature, the flux from the jets should be 10 times weaker in the 2--5~keV range and 60 times weaker in the 8--20~keV range than during the phase when the observer can see the funnel to  maximum depth. The contribution of the jets to the observed X-ray flux should therefore be negligibly small for the 'edge-on disc'. The fact that the observer does not see the jets during this precession phase is confirmed by the absence of a correlation between the soft and hard 
emission (Figs.~\ref{fig:xray+xray}d,\ref{fig:xray+xray2}d). 

\cite{Cherepashchuk2005}, based on the RXTE/ASM and Ginga data, and also \cite{Filippova2006}, based on the RXTE/PCA data, found that during the `edge-on' precession phase the flux in the standard X-ray range is only 2 times smaller than at the phase of maximal opening of the funnel. We confirm this result: despite the fact that the jet and the inner parts of the funnel are obscured during the `edge-on' phase, the observer sees a considerable X-ray flux. We suggest that the radiation possibly comes out of the funnel and is then reflected from the dispersed clouds of gas located above the funnel. 

It is considered that the main portion of the bolometric luminosity of SS\,433 ($~\sim10^{40}$~erg/s) is emitted in the inner parts of the funnel of the supercritical accretion disc in the X-ray range \citep{Fabrika2004}. The gas of the jet and the wind of the supercritical accretion disc, irradiated by the inner parts of the funnel, will scatter this emission. Indeed, even a small optical depth of $\tau\sim10^{-4}$ in the dispersed clouds is enough to see a considerable X-ray flux at any precession phase. The gas of the jets and the sparse gas possibly filling the funnel could act as a scattering screen directly above the funnel ($\sim10^{12}-10^{13}$~cm). At the distance $\gtrsim 10^{14}$, the jets interact with the wind of the supercritical accretion disc because of their precession. This interaction is considered to be a compulsory requirement for the jets to be able to emit in HI and HeI lines in the optical range at the distances $\sim 10^{14}-3\times 10^{15}$~cm \citep{Panferov1997}. As a result of 
this interaction between the jets and the wind, extensive gaseous structures may also form, which will scatter X-rays coming from the inner parts of the accretion disc. 

\begin{figure}
  \includegraphics[width=0.5\textwidth]{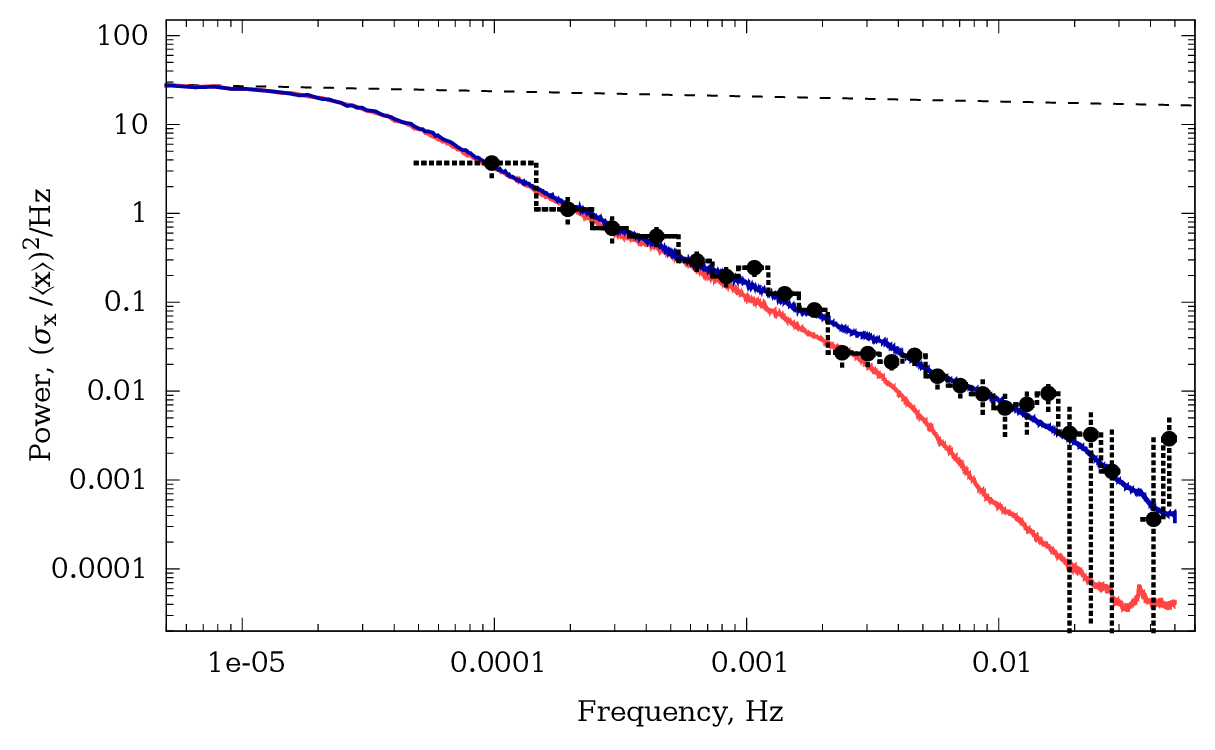}
\caption{Power spectra corresponding to the `edge-on' orientation. The circles~--- observed power spectrum (same as in Fig.~\ref{fig:pds_data}), grey (red) line~--- 'reflecting clouds' model, {\bf dark grey (blue)} line~--- `reflecting clouds and jet' model. The initial power spectrum of the incident funnel radiation $P\propto f^{-0.06}$ is shown by the dashed line. (A colour version of this figure is available in the online journal.)}
\label{fig:pds_crossover}
\end{figure}

Reflection from these gas structures (clouds), like in the case of reflection on the funnel walls in Sec.~\ref{sec:funnel_model}, should smear out the variability of the incident radiation and change its power spectrum. The original PDS of the inner parts of the funnel, as we suggested in Sec.~\ref{sec:funnel_model}, is flat (we will discuss this in more detail in the next section). The observed power spectrum would be determined by the configuration of clouds, the region extent, the distribution of clouds by size, etc. We performed the modelling and determined that under certain conditions, such a cloud system may well yield a power spectrum with a constant slope in the observed frequency range (Figs.~\ref{fig:pds_data},\ref{fig:pds_crossover}).

In our model we assumed that the clouds fill a cone with an opening angle corresponding to that of the funnel (50\degr). The number of clouds in the unit volume decreases with distance as $\propto r^{-2}$. All the clouds are located at distances ranging from $r_{min}$ to $r_{max}$ from the center of the disc; moreover, $r_{min}\gtrsim l_f$, because the deeper regions should be obscured by the funnel wall. The clouds have different sizes. The distribution of clouds by size obeys a power law: 
\begin{equation}
N=N_0\left(d/d_{min}\right)^{-s}
\label{eq:cloud_distr}
\end{equation}
where $d_{min}$ is the minimum size of a cloud. Obviously, $d_{min} < r_{min}$. We adopted $d_{min}=r_{min}/3$. For the maximum cloud size we adopted the criterion $d_{max}=r_{max}/10$. We will refer to this model component as `reflecting clouds' model.

The second component of the model is the `reflecting jet'. The jet is represented by a cylinder with the diameter $l_j$, filled uniformly by clouds. 
The clouds of the jet may overlap each other, and therefore the flux of the illuminating radiation will undergo absorption:
\[
F(r)\propto {1\over r^2}e^{-r/r_\sigma},
\]
$r_\sigma=(n\sigma)^{-1}$~--- mean free path of a photon, n~--- jet gas density, $\sigma$~--- electron scattering cross section. 

For this model we used the Monte-Carlo method to compute the response functions; we then used them to construct the model power spectra as described in Sec.~\ref{sec:funnel_model}. The 'clouds+jet' model PDS is shown in Fig.~\ref{fig:pds_crossover}. For comparison, we show the PDS of the first component~--- the reflecting clouds. At the frequency $\approx2\times10^{-5}$~Hz, approximately corresponding to the light-crossing time $r_{max}/c$, both power spectra exhibit a break. We adopted $r_{max}=5\times 10^{14}$~cm in the  figure. This distance corresponds to 0.7 days of jet propagation. Starting from this distance (and further on), optical line emission appears in the jets \citep{BorisFabr1987,Vermeulen1993}. The optical emission of the jets of SS\,433 is considered to emerge due to the interaction of dense jet clouds with the gas of the wind of the supercritical disc. There is no smearing out of variability at the frequencies below the break, and the PDS has a slope of 0.06, which corresponds to the PDS of 
the inner parts of the funnel. At the frequencies above the break, the slope is determined by the exponent $s$ in the distribution (\ref{eq:cloud_distr}). The slope in Fig.~\ref{fig:pds_crossover} corresponds to the observed one for $s \approx 2.5$.

At the frequency $\approx 2\times10^{-3}$~Hz, the `reflecting clouds' model exhibits a second break, determined by the minimum size $r_{min}=3\times10^{12}$~cm in Fig.~\ref{fig:pds_crossover}. At $r_{min}\sim 10^{10}$~cm the frequency of this break falls into the $f>0.1$~Hz range. However, the distance $r_{min}$ can not be less then the funnel size.

In the `clouds+jet' model, the dip above the break at $2\times10^{-3}$~Hz is compensated by the contribution from the jets. Below the break frequency, the jet contributes practically nothing to the power spectrum, because the funnel radiation does not reach the $r \gg r_\sigma$ regions due to scattering. The model power spectra in Fig.~\ref{fig:pds_crossover} are obtained for the value $r_\sigma\sim10^{12}$~cm (jet diameter $l_j\sim 2\times 10^{11}$~cm), which agrees with the expected density of the jets. In Sec.~\ref{sec:jet_model} we found that the density of fraction C, which has a biggest contribution to the jet mass, is $5\times10^{11}$~cm. For the Thomson cross section, the mean free path of a photon of $r_\sigma \sim 3\times 10^{12}$~cm corresponds to this density. 

Thus, the power spectrum corresponding to the 'edge-on' orientation of the system has neither a flat portion, nor a break. Nonetheless, as we were able to show, this specific form does not contradict the idea that at this precession phase, the observer also sees the emission that forms in the inner parts of the supercritical accretion disc. Reflection from different gaseous structures (the sparse gas filling the funnel, the jets themselves, and also the gas clouds that emerge in the region of interaction between the jets and the wind) may well ensure the power spectrum observed during the 'edge-on' precession phases. 

\subsection{The flat PDS and viscous time-scale}
\label{sec:vis_time}
We have found that at the precession phases of open disc the PDS of SS\,433 in the $10^{-4}$--$10^{-3}$~Hz frequency range is about flat ($\beta\approx0.06$). Both in the funnel model and in the jet model we can explain only the position of the break in the PDS and the shape of the spectrum at the frequencies above the break. In all our models, including the 'reflecting clouds' model discussed in the previous section, the PDS with $\beta=0.06$ has been adopted as that in observed spectrum (Fig.~\ref{fig:pds_data},\ref{fig:pds_data_g1g4}a). We believe that the flat spectrum corresponds to the original power spectrum of the innermost parts of the supercritical accretion disc, which are inaccessible for direct observation, and  may be an intrinsic property of the supercritical accretion. The mechanism which produces a flat power spectrum may control both the emerging radiation and the jet formation.

The presence of the flat region in the SS\,433 PDS is discrepant with the result of \cite{Revn2006}, who obtained the slope of $1.5$ in a wide frequency interval from $10^{-7}$ to $10^{-2}$~Hz. However, these authors had reliable and homogeneous data only for frequencies below $10^{-5}$~Hz. At higher frequencies, their power spectra were constructed based on the EXOSAT/ME data, which are more noisy than the RXTE/PCA data used by us.

The formation of a power-law spectrum in SS\,433 is related to the fluctuations of viscosity in the accretion disc \citep{Revn2006}. The mechanism of the formation of power-law PDS due to random fluctuations of the viscosity parameter $\alpha$ was described in detail by~\cite{Lyubarskii1997}. It is assumed that the fluctuations are small and occur independently at different disc radii. Their typical time-scale should be of the order of the viscous time, or a time of matter propagation through the disc \citep{ShakSun1973}:
\begin{equation}
\label{eq:time_visc}
t_a(r)=\left[\alpha\left(h\over r\right)^2 \Omega_K \right]^{-1}
\end{equation}
where $h(r)$ is the thickness of the disc at a given radius, and $\Omega_K$ is the Keplerian angular velocity. At a larger time-scales the fluctuations of $\alpha$ become independent at each radius. The further from the disc center, the greater the typical time-scale of the fluctuations. 

Viscosity fluctuations in a ring of radius $r$ should in turn lead to perturbations of the accretion rate in the given ring. As matter passes through the disc, these perturbations accumulate, i.e., at every radius, the disc 'remembers' all the fluctuations that occurred in the outer parts of the disc. Therefore, when matter reaches the inner disc radius, where the maximum energy release takes place, it bears information about the perturbations that occurred at every radius and time-scale. \cite{Lyubarskii1997} showed that in this case, a power-law PDS should be observed, the slope of which should be determined by the amplitude of the variations of $\alpha$ at different radii. In particular, if the variation amplitude is the same at all radii, the power spectrum should be $P\propto f^{-1}$.  

A continuous power-law power spectrum should be observed only if all the terms in
equation~(\ref{eq:time_visc}) vary smoothly with radius. In the case of a supercritical accretion disc, this is not so. According to \cite{ShakSun1973}, a supercritical disc should possess a special radius $r_{sp}$~--- the spherization radius, below which the supercritical-disc properties begin to manifest themselves, and the structure of the disc changes abruptly.
\begin{equation}
\label{eq:r_sp}
r_{sp}\sim r_{in}{\dot{M}_0\over \dot{M}_{Edd}},
\end{equation}
where $\dot{M}_{Edd}$ is the Eddington accretion rate and $r_{in}$ is the inner disc radius. At distances $r < r_{sp}$, the radiation pressure is greater than gravity. The disc becomes geometrically thick with $h/r \sim 1$. Below the spherization radius the disc loses matter as a wind, in such a way that it remains locally Eddington in every point: $\dot{M}(r)=\dot{M}_0(r/r_{sp})$. At $r > r_{sp}$ region the disc is considered to be standard. Its thickness is relatively small: $h/r\sim 0.03-0.1$ \citep{ShakSun1973}. Thus, the $h/r$ ratio, and therefore, the characteristic viscous time $t_a(r)$, should undergo a discontinuity at the spherization radius.

Due to the abrupt change of the $h/r$ value $t_a$ at the spherization radius will have a discontinuity of at least two orders of magnitude. This should affect the PDS. If we insert into (\ref{eq:time_visc}) 
the Keplerian velocity at $r_{sp}$ we can estimate the typical frequencies at this radius (Hz):
\begin{equation}
\label{eq:f_sp}
f(r_{sp})=t_a^{-1}(r_{sp})\approx 3\times 10^{-2}\alpha_{0.1}\left({h\over r}\right)^2 m_{10}^{-1} \dot{m}_{300}^{-3/2},
\end{equation}
where $\alpha$ is in the units of 0.1, black-hole mass is in $10M_{\odot}$, and the accretion rate is in the units of $300\dot{M}_{Edd}$. 

\begin{figure}
  \includegraphics[width=0.4\textwidth]{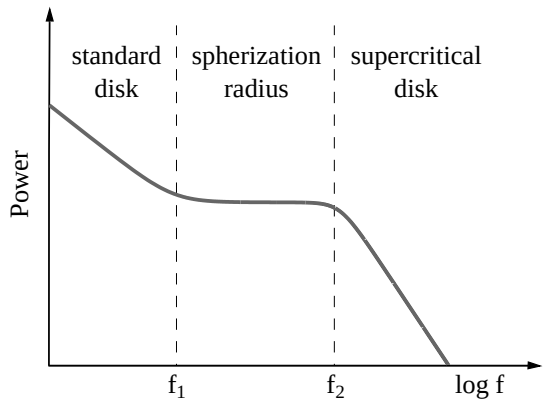}
\caption{A probable power spectrum of the supercritical accretion disc. The $f_1$ frequency corresponds to the viscous times immediately to spherization radius; $f_2$ is a frequency at the spherization radius.} 
\label{fig:pds_scheme}
\end{figure}

We believe that the power spectrum of the supercritical disc may have the following structure (Fig.~\ref{fig:pds_scheme}). The frequencies $f_1$ and $f_2$ will correspond to the values $h/r\sim 0.03-0.1$ and $h/r\sim 1$ in equation (\ref{eq:f_sp}): immediately to the $r_{sp}$ and at the $r_{sp}$ respectively. (i)~In the $<f_1$ frequency range, a power law should be observed, the exponent of which is determined by the fluctuations of the standard-disc viscosity. (ii)~At the frequencies between $f_1$ and $f_2$, the shape of the PDS is determined by the fluctuations of $\alpha$ at $r_{sp}$ itself. If the viscosity at the $r_{sp}$ varies in a random and independent manner, as assumed for all the orbits \citep{Lyubarskii1997}, the PDS at this radius should be quasi-flat. (iii)~In the $>f_2$ frequency range the power spectrum of the disc may once again be power-law-like. Here the shape of the PDS will be determined by the processes taking place in the supercritical region of the disc. All information about changes 
in the accretion rate (the $\alpha$ parameter) may be transferred inwards from the spherization radius with almost free-fall time. In addition, the viscosity ($\alpha$) in the supercritical region may differ from that in the standard disc. Direct numerical simulations could answer this question. 

With this idea, we can estimate the extent of the quasi-flat region of the supercritical-disc PDS. It amounts to $f_2/f_1\sim 10^{2-3}$ (depending of the value of $h/r$ in the standard disc), i.e., about 2-3 orders of magnitude in frequency. The extent of the flat region does not depend on accretion rate if $\dot{M}_0 \gg \dot{M}_{Edd}$, but can depend on the variations of the $\alpha$ parameter, which we set to be constant in our estimation. 

For SS\,433, we can adopt the mass of the black hole $M=10M_{\odot}$, and the accretion rate $\dot{M}_0=300 \dot{M}_{Edd}$ \citep{Fabrika2004}. In this case, $r_{sp}\sim 3\times 10^{9}$~cm, and the frequencies are $f_1\approx 3\times10^{-5}$~Hz and $f_2\approx 3\times 10^{-2}$~Hz. The flat region in Fig.~\ref{fig:pds_data} ($10^{-4}-10^{-3}$~Hz) falls within that frequency range. We do not have observations long enough to study the PDS at the frequencies below $10^{-4}$~Hz. Nonetheless, we can expect that the flat part will come to an end in the $3\times 10^{-5} - 10^{-4}$~Hz region, because the PDS with $\beta\approx 1.5$ has been already found by \cite{Revn2006}.

An intrinsic PDS of the innermost parts of SS\,433 supercritical disc may have the shape described above. However this emission is inaccessible for direct observation. The observer sees the emission reflected from the funnel walls, and the intrinsic jet emission. These processes distort the original PDS of SS\,433 (Sec.~\ref{sec:funnel_model}). In particular, due to both the scattering in the funnel and the jet cooling, a break appears at $2\times10^{-3}$~Hz (Figs.~\ref{fig:pds_data},\ref{fig:pds_results}). Therefore, it is impossible to study the shape, slope, and breaks at $10^{-2}-10^{-1}$~Hz in an undistorted PDS of the supercritical disc of SS\,433. However, we believe that the power spectra of other objects with supercritical accretion (ULXs~--- ultraluminous X-ray sources are the most likely candidates), where the observer sees the innermost regions of the funnel \citep{FabrMesch2001,Poutanen2007}, should have extensive flat regions, spreading over 2-3 orders of magnitude in frequency.

\section{Conclusions}
We analyzed a rapid X-ray variability of SS\,433 based on the RXTE data in the $10^{-4} - 5\times 10^{-2}$\,Hz frequency range. The shape of the cross-correlation functions and power spectra depends drastically on the system's precession phase. At the phase corresponding to the maximal opening of the funnel of the supercritical disc, the PDS has a quasi-flat region with a power-law index of $0.06$; at the frequency $1.7\times10^{-3}$~Hz, a break appears, above which the PDS has an index of $1.67$. The PDS corresponding to the `edge-on' phase, when the observer cannot see the funnel, exhibits neither a flat region nor a break. A single power-law with the index $1.34$ is observed at all frequencies.

The CCFs of the X-ray emission in the 2--5\,keV and 8--20\,keV ranges have a complex profile with a pronounced asymmetry indicating a lag of the soft emission behind the hard emission at times ranging from several to 250 seconds. Gaussian analysis allows to distinguish three components in the profile, with FWHM of 20, 50, and 150~s and typical delays of 5, 25, and 45 s, respectively. This effect is most conspicuous at maximal opening of the funnel, and it disappears at the `edge-on' orientations. The data of simultaneous RXTE and optical observations has shown that the soft X-rays also lag behind the optical, but the hard X-rays reach the observer simultaneously with the optical. 

We conclude that the X-ray and optical variability of SS\,433 is fully determined by the visibility of the funnel in the supercritical accretion disc. The delay of soft X-ray emission relative to the hard and optical emission is consistent with the idea \citep{Revn2004,MedvFabr2010} that the soft emission is generated mostly in the jets but the hard emission is reflected at the outer funnel walls and the optical emission is a result of reprocession of the hard X-rays.

We investigated two mechanisms that may explain the observed variability of SS\,433, 1) reflection and scattering of emission in the wind funnel of the supercritical disc (the funnel model, applicable both to X-ray and optical emission) and 2) the cooling of the jet gas (jet model, applicable only to X-ray emission). Both models well reproduce the position of the break and the slope at high frequencies in the X-ray power spectrum corresponding to the phase of maximal opening of the funnel. Optical emission of SS\,433 cannot be formed in the jets, and therefore optical variability is determined fully by the funnel. 

Modelling of optical PDS yields the following estimates of the funnel parameters: the funnel wall length $l_f\sim(3-4.5)\times10^{12}$~cm, and opening angle $\vartheta_f\sim 50^\circ$. Modelling of X-ray power spectra results in similar quantities. We believe that unlike the optical, the X-ray variability is determined by the contribution of both mechanisms (funnel and jet); moreover, the contribution of the jets to the X-ray variability is significantly greater. It also follows from a comparison of the variability amplitudes of X-rays and optical PDS that the jets are a more variable source than the funnel.

Our modelling of X-ray CCFs and PDS in the cooling X-ray jet model showed that the jets may consist of three fractions of clouds with different densities: $8\times10^{13}$, $3\times10^{13}$ and $5\times10^{11}$ cm$^{-3}$, and most of the jet mass is concentrated in the last fraction. 

In the `edge-on' precession phase, when both the jets and the funnel are obscured by the wind, the observer sees a significant X-ray flux. The unobscured, cooled outer parts of the jets cannot provide the observed X-ray flux. It was suggested that the clouds of the jet and the wind, which `see' the funnel directly, should scatter a part of the funnel's emission. We have found that the PDS observed at the `edge-on' can indeed be reproduced if the emission is scattered on the extensive gaseous structures located above the funnel at distances up to $\sim10^{15}$~cm. 

At the precession phase when the funnel is maximally open to the observer, the PDS of SS\,433 exhibit a flat part in the $10^{-4} - 2\times 10^{-3}$\,Hz frequency range. We argue that the presence of such a part is related to the abrupt change in the disc structure and the viscous time at the spherization radius. In this place the accretion disc becomes thick $h/r \sim 1$, which reduces drastically the time of passage of matter through the disc. The position of the flat part in the PDS depends on the mass of the black hole, accretion rate and viscosity in the disc, however, its extent (2-3 orders of magnitude in frequency) depends only on the change in viscosity between subcritical and supercritical regions. Simple estimates of the position of the flat part in SS\,433 yield a $3\times 10^{-5} - 3\times 10^{-2}$~Hz interval. The observed flat portion in SS\,433 really is located within this interval. However, as we have seen, the funnel and the cooling jets smooth out the variability at high frequencies; a 
break at $1.7\times10^{-3}$~Hz is observed in the PDS. At the frequencies $ < 3\times 10^{-5}$~Hz, the power spectrum should again become power-law-like (this was shown by \citet{Revn2006}). 

We cannot see the whole flat part in SS\,433, and therefore, cannot analyze the higher frequencies which would be determined by the flow of matter below the spherization radius. However one may expect that flat parts should be observed in other supercritical accretion discs, and in particular, in ultraluminous X-ray sources. 

{\bf Acknowledgements}~~

The authors are grateful to Y.\,Lyubarskii, V.\,Goranskij and anonymous referee for their helpful comments. A.\,M. is supported by the Saint-Petersburg State University through a research grant 6.50.1555.2013. S.\,F. 
acknowledges support of the Russian Government Program of Competitive Growth of Kazan Federal University. A.\,V. was partly support by grants of the Russian Foundation for Basic Research (projects no. 13-02-00885, 12-02-31548 and 14-02-00759) and the Grants of the President of the Russian Federation for Support of Young Russian Scientists (MK-6686.2013.2).
The research was supported by the Program for Leading Scientific Schools of Russia N\,2043.2014.2.

\appendix
\section{Non-expanding radiative blobs}
In the case of the non-expanding blob model the first law of thermodynamics can be written as:
\begin{equation}
\delta q = d\varepsilon,
\end{equation}
$\varepsilon= (3/2)n \theta$~--- internal energy of 1 cm$^3$ of an ideal gas. The energy losses occur only due to emission; for the temperature in keV (erg cm$^{-3}$ s$^{-1}$):
\begin{equation}
\frac{\delta q}{dt}= - n^2\int\limits_0^\infty J_\nu(\theta) d\nu \approx -6.26\times 10^{-24}n^2  \sqrt{\theta}.
\end{equation} 
We obtain the differential equation:
\begin{equation}
\frac{d\theta}{dt} = - n\alpha \sqrt{\theta},
\end{equation}
where $\alpha\approx2.6\times 10^{-15}$ (keV$^{1/2}$ cm$^3$ s$^{-1}$). The physical solution to the equation would be the descending branch of the parabola
\begin{equation}
\label{eq:theta_rad}
\theta(t)=\theta_0\left(1-\frac{t}{t_{cool}}\right)^2,
\end{equation}
where $t_{cool}=(2\sqrt{\theta_0})/(\alpha n)\approx 7.7\times 10^{14} \sqrt{\theta_0}/n$ (s) is the time in which the blob loses all thermal energy. 

We can also introduce the typical cooling time, during which the temperature drops $e$-fold. Substituting $\theta(t)=\theta_0/e$ into equation (\ref{eq:theta_rad}) and solving it, we obtain
\begin{equation}
\label{eq:te_rad}
t_e^{rad}=t_{cool}(1-e^{-1/2})\approx 120 \sqrt{\theta_{17}}/n_{13} (s),
\end{equation}
where $\theta_{17}$ is in the units of 17~keV, $n_{13}$~--- in the units of $10^{13}$~cm$^{-3}$.

\section{Adiabatic blobs}
In the adiabatic model the cooling takes place only due to expansion. In this case, the first law of thermodynamics for 1 cm$^3$ of gas has the following form:
\begin{equation}
d\varepsilon + p n d\left(\frac1{n}\right) = 0 
\end{equation}
Inserting the expressions for the internal energy $\varepsilon= (3/2)n \theta$ and density $p=n\theta$, we obtain the differential equation: 
\begin{equation}
\frac{d\theta}{\theta} = \frac23\frac{dn}{n}.
\end{equation}
Its solution expresses the dependence of temperature on density for the adiabatic process:
\begin{equation}
\theta=\theta_0 \left(\frac{n}{n_0}\right)^{2/3}.
\end{equation}

The increase of the blob size $r_b$ occurs with the speed of sound
\begin{equation}
r_b(t) = r_{b0} + c_s(t) t,
\end{equation}
$r_{b0}$~--- a size of the blobs at the initial moment of time.
\begin{equation}
c_s^2 = \frac{1}{\mu m_p} \frac{dp}{dn} = \frac{5\theta_0}{3 \mu m_p}\left(\frac{n}{n_0}\right)^{2/3},
\end{equation}
or
\begin{equation}
c_s = c_{s0} \left(\frac{n}{n_0}\right)^{1/3}.
\end{equation}
The initial speed of sound is equal to $c_{s0}\approx5.1\times10^7 \sqrt{\theta_0}$cm~s$^{-1}$ (for the temperature in keV). The density and size of the blobs are connected by the ratio:
\begin{equation}
n(t) = \frac1{V}\frac{M_b}{\mu m_p} = \frac3{4\pi r_b^3(t)}\frac{M_b}{\mu m_p}, 
\end{equation}
Inserting into this formula the expression for $r_b(t)$, we obtain the equation
\begin{equation}
n(t) = \frac{3}{4\pi r_{b0}^3}\frac{M_b}{\mu m_p}\left[1 + \frac{c_{s0}}{r_{b0}}\left(\frac{n(t)}{n_0}\right)^{1/3} t\right]^{-3}.
\end{equation}
Let us introduce the typical expansion time $t_{exp}=r_{b0}/c_{s0}$ and the dimensionless time $\tau=t/t_{exp}$. The equation can be reduced to a quadratic equation with respect to $(n/n_0)^{1/3}$:
\begin{equation}
\left(\frac{n}{n_o}\right)^{1/3} \left(1+\tau \left(\frac{n}{n_o}\right)^{1/3} \right)=1.
\end{equation}
The physical solution here is 
\begin{equation}
n(\tau)=n_0 \left(\frac{\sqrt{1+4\tau}-1}{2\tau}\right)^3.
\end{equation}
From it, we can easily derive the expressions for the remaining blob parameters: 
\begin{equation}
\theta(\tau)=\theta_0 \left(\frac{\sqrt{1+4\tau}-1}{2\tau}\right)^2,
\end{equation}
\begin{equation}
r_b(\tau)=r_{b0}\left(1+\frac{\sqrt{1+4\tau}-1}{2}\right).
\end{equation}
Typical cooling time:
\begin{equation}
t_e^{ad}=t_{exp} (e-e^{1/2})\approx 2.1\times10^{-8}r_{b0} \theta_0^{-1/2} 
\end{equation}
Taking into account the relation (equation~\ref{eq:blobsize}) between the density of the blobs and their size, we obtain
\begin{equation}
\label{eq:te_ad}
t_e^{ad}\approx 4.1\times10^6 n_0^{-1/3} \theta_0^{-1/2} \approx 46 n_{13}^{-1/3} \theta_{17}^{-1/2} (s),
\end{equation}
where $\theta_{17}$ is expressed in the units of 17~keV, $n_{13}$~--- in the units of $10^{13}$~cm$^{-3}$.
\end{document}